\documentclass[manuscript,screen,authorversion]{acmart}

\usepackage{caption}

\usepackage{subcaption}
\usepackage{tabularx}
\usepackage{graphicx}

\usepackage{siunitx}
\usepackage{tikz}
\usepackage{pifont}% http://ctan.org/pkg/pifont

\usepackage{diagbox}
\newcommand{\cmark}{\ding{51}}%
\newcommand{\xmark}{\ding{55}}%
\newcolumntype{L}{>{\raggedright\arraybackslash}X}

\AtBeginDocument{%
  \providecommand\BibTeX{{%
    \normalfont B\kern-0.5em{\scshape i\kern-0.25em b}\kern-0.8em\TeX}}}

\begin{document}

%%
%% The "title" command has an optional parameter,
%% allowing the author to define a "short title" to be used in page headers.
\title[Rethinking Evaluation in Human--AI Collaborative Design]{From Metrics to Meaning: Time to Rethink Evaluation in Human--AI Collaborative Design}

%%
%% The "author" command and its associated commands are used to define
%% the authors and their affiliations.
%% Of note is the shared affiliation of the first two authors, and the
%% "authornote" and "authornotemark" commands
%% used to denote shared contribution to the research.
%\author{Ben Trovato}
%\authornote{Both authors contributed equally to this research.}
%\email{trovato@corporation.com}
%\orcid{1234-5678-9012}
%\author{G.K.M. Tobin}
%\authornotemark[1]
%\email{webmaster@marysville-ohio.com}
%\affiliation{%
%  \institution{Institute for Clarity in Documentation}
%  \streetaddress{P.O. Box 1212}
%  \city{Dublin}
%  \state{Ohio}
%  \country{USA}
%  \postcode{43017-6221}
%}
\author{Sean~P.~Walton}
\email{s.p.walton@swansea.ac.uk}
\affiliation{
  \institution{Computational Foundry, Faculty of Science and Engineering, Swansea University}
 \streetaddress{Swansea University}
  \city{Swansea}
  \country{Wales, UK}
}

\author{Ben~J.~Evans}
\affiliation{
  \institution{Department of Aerospace Engineering, Faculty of Science and Engineering, Swansea University}
 \streetaddress{Swansea University}
  \city{Swansea}
  \country{Wales, UK}
}

\author{Alma~A.~M.~Rahat}
\affiliation{
  \institution{Computational Foundry, Faculty of Science and Engineering, Swansea University}
 \streetaddress{Swansea University}
  \city{Swansea}
  \country{Wales, UK}
}
\author{James~Stovold}
\affiliation{
  \institution{Lancaster University Leipzig}
  \country{Germany}
}
\author{Jakub~Vincalek}
\affiliation{
  \institution{Computational Foundry, Faculty of Science and Engineering, Swansea University}
 \streetaddress{Swansea University}
  \city{Swansea}
  \country{Wales, UK}
}

%%
%% By default, the full list of authors will be used in the page
%% headers. Often, this list is too long, and will overlap
%% other information printed in the page headers. This command allows
%% the author to define a more concise list
%% of authors' names for this purpose.
\renewcommand{\shortauthors}{Walton, et al.}

%%
%% The abstract is a short summary of the work to be presented in the
%% article.
\begin{abstract}
As AI systems increasingly shape decision making in creative design contexts, understanding how humans engage with these tools has become a critical challenge for interactive intelligent systems research. This paper contributes a challenge to rethink how to evaluate human--AI collaborative systems, advocating for a more nuanced and multidimensional approach. Findings from one of the largest field studies to date (n = 808) of a human--AI co-creative system, The Genetic Car Designer, complemented by a controlled lab study (n = 12) are presented. The system is based on an interactive evolutionary algorithm where participants were tasked with designing a simple two dimensional representation of a car. Participants were exposed to galleries of design suggestions generated by an intelligent system, MAP--Elites, and a random control. Results indicate that exposure to galleries generated by MAP--Elites significantly enhanced both cognitive and behavioural engagement, leading to higher-quality design outcomes. Crucially for the wider community, the analysis reveals that conventional evaluation methods, which often focus on solely behavioural and design quality metrics, fail to capture the full spectrum of user engagement. By considering the human--AI design process as a changing emotional, behavioural and cognitive state of the designer, we propose evaluating human--AI systems holistically and considering intelligent systems as a core part of the user experience---not simply a back end tool.
\end{abstract}

%%
%% The code below is generated by the tool at http://dl.acm.org/ccs.cfm.
%% Please copy and paste the code instead of the example below.
%%

\begin{CCSXML}
<ccs2012>
   <concept>
       <concept_id>10003120.10003123.10011759</concept_id>
       <concept_desc>Human-centered computing~Empirical studies in interaction design</concept_desc>
       <concept_significance>500</concept_significance>
       </concept>
   <concept>
       <concept_id>10010147.10010178</concept_id>
       <concept_desc>Computing methodologies~Artificial intelligence</concept_desc>
       <concept_significance>500</concept_significance>
       </concept>
   <concept>
       <concept_id>10010405.10010432.10010439.10010440</concept_id>
       <concept_desc>Applied computing~Computer-aided design</concept_desc>
       <concept_significance>500</concept_significance>
       </concept>
 </ccs2012>
\end{CCSXML}

\ccsdesc[500]{Human-centered computing~Empirical studies in interaction design}
\ccsdesc[500]{Computing methodologies~Artificial intelligence}
\ccsdesc[500]{Applied computing~Computer-aided design}

%%
%% Keywords. The author(s) should pick words that accurately describe
%% the work being presented. Separate the keywords with commas.
\keywords{Mixed initiative, MAP--Elites, Human--AI Collaboration, Optimisation, Design, User Evaluation}

\received{20 February 2007}
\received[revised]{12 March 2009}
\received[accepted]{5 June 2009}

%%
%% This command processes the author and affiliation and title
%% information and builds the first part of the formatted document.
\maketitle

\section{Introduction}
Despite increasing sophistication of automatic design algorithms there is limited uptake of state-of-the-art techniques within the engineering industry. When asked the reason for this limited uptake, engineers report a lack of trust towards automated design algorithms along with a strong desire to remain a core part of the design loop~\cite{Vincalek2021-lz2, Vincalek2021-lz}. Mixed-initiative tools which aim to assist humans and computers to collaborate during design tasks~\cite{Charity2022-bl} are a natural solution to the trust issue expressed by engineers. 

Various fields have approached the problem of human--AI collaborative, or mixed-initiative, design independently. Interestingly, the two domains which arguably have contributed the most to this area, HCI and video games research, have approached the problem from opposite directions. In video games research mixed-initiative tools began as procedural content generation tools (PCG) designed to replace human designers and fully automate design tasks~\cite{Melotti2019-rt, Ruela2018-ah, Baldwin2017-me}. Over time the PCG community shifted emphasis onto how generative tools could be used to support the design process~\cite{Cook2017-gh} leading to a wide range of mixed-initiative algorithms and techniques~\cite{Liapis2013-at, Baldwin2017-me, Ruela2018-ah, Von_Rymon_Lipinski2019-tk, Charity2022-bl, Earle2022-pk}. In contrast, research into mixed-initiative systems within the HCI community began by considering how to build tools to support designers as part of the creative process. These tools often involved new methods of presenting existing human crafted designs as useful examples~\cite{Lee2010-wa, Ngoon2018-iu}. More recently, the proliferation of generative AI has led to a number of researchers exploring the use of algorithms to generate examples in tools which support creativity, instead of using existing human crafted designs, resulting in mixed-initiative systems~\cite{Swearngin2020-xt, Louie2020-gl}. Although there are a range of approaches to building mixed-initiative design tools, there continues to be a significant gap in our understanding of how these tools impact the design process and experience of designers~\cite{Craveirinha2015-hm, Chan2022-gh}. This knowledge gap makes improving upon, and therefore adapting, techniques to new contexts challenging~\cite{Lu2017-ll} and is the core focus of our work herein.

The key contributions of this work are:
\begin{itemize}
    \item Rigorous evaluation of a human--AI collaborative design tool based on galleries of design suggestions. This evaluation is based on (to our knowledge) one of the largest field studies of a mixed-initiative tool (n=808) complemented by a smaller controlled lab study (n=12).
    \item Evidence that exposure to galleries of examples generated by AI leads to increased engagement in the design process and this leads to better quality design outcomes.
    \item Findings that the value of using a MAP--Elites algorithm to generate example designs changes based on the approach of the designer. Furthermore, a designers' approach may change throughout the process, which opens up a potential avenue for research into human--AI collaborative environments which adapt to user preference.
    \item A challenge to rethink how we evaluate human--AI collaborative environments. Specifically, we found that the act of simply viewing a suggested design has an influence on the design process, and therefore current evaluation methodologies which focus on quantitative measures of how often users edit or copy suggestions do not tell the full story.
\end{itemize}

%Use R1 comments to reframe and strengthen

%Address the concern "In this fast-moving Gen AI era, my major concern of this submission falls into its lack of novelty and contributions, especially in design quality/complexity and the method."

\section{Related Work}
%part of what makes this research compelling is that it seeks to bridge the gap between techniques that have been effectively applied in game design and the space of engineering design. It follows that this paper would be well-served by being grounded in well-established principles from this domain.

%Add references from R2 - comment on size of studies

%Where does engineering design fit

\subsection{Galleries of Examples Support Creativity}
It is well understood that designers use examples for inspiration, and there is a long history within HCI of developing systems to aid designers to explore galleries of existing examples. For example, Lee et al.~\cite{Lee2010-wa} designed the Adaptive Ideas web design tool which, based on the elements the designer is interested in, presents a gallery of designs from a pre-selected set of existing web applications. The examples presented in the galleries were selected based on several metrics adapting to the current state of the user's design. User studies (n=30) showed that the Adaptive Ideas web design tool improved both the quality of designs and the experience of the designers.

In their work Ngoon et al.~\cite{Ngoon2018-iu} designed a system, CritiqueKit to support reviewers in providing useful feedback, which in itself is a creative task. As reviewers wrote feedback, the system used a text classifier to indicate in real time what aspects of effective feedback was missing from the in-progress review. In addition the system offered suggestions of previously provided feedback for reuse in a list; it was this adaptive set of suggestions which had the most impact on helping reviewers in their task during a user study (n=8). 

Another example of a tool which uses a gallery of suggestions is GUIComp~\cite{Lee2020-db}. Designed to be a companion for GUI design a key element of GUIComp is a recommendation panel which is a subset of a large corpus of screenshots of existing GUIs presented as a gallery. The authors of GUIComp made the decision to firstly present a number of examples which are identified as being similar to the current state of the users' design, along with some random examples to increase diversity in design choices. Users were also enabled to "keep" some of the recommendations as templates, pinning them into the gallery. In their user study (n=30) it was found that the recommendation panel was useful for inspiration at the beginning of the design process but because less useful later on, mirroring the findings of Duan et al.~\cite{Duan2024-bi} with respect to generated feedback. They also found that users desired explanations of why recommendations were suggested to them.

\subsection{Galleries of Examples can be Generated by Algorithms}
In contrast to CritiqueKit and GUIComp, Scout~\cite{Swearngin2020-xt} does not use existing designs as suggestions but instead generates alternative designs based on constraints defined by the designer. Using a quality metric designed by the developers the highest quality designs generated are presented to the designer in a gallery. Through their user study (n=18) they found that Scout can aid designers in producing designs they otherwise would not have thought of and can help them avoid the fixation effect where designers focus in on an early design. 

Louie et al.~\cite{Louie2020-gl} took an alternative approach and used a gallery of examples to facilitate the human giving feedback to the AI. They developed tools to enable composers to influence the music generated by deep neural networks. These AI-steering tools included example based sliders whereby composers could control the similarity of the generated music to pre-existing examples. In their user study (n=21) they found that the steering tools increased the sense of collaboration with the AI and contributed to a greater sense of ownership over the composition relative to the AI. 

In video games research approaches where designers are presented galleries of suggestions or recommendations are referred to as \emph{mutant shopping} approaches~\cite{Lai2022-qv}. This is in reference to the evolutionary algorithms which often drive the generation of designs. For example, Alvarez et al. \cite{Alvarez2018-am} use a mutant shopping approach in their evolutionary dungeon designer. An underlying evolutionary algorithm generates a collection of candidate dungeons based on computer calculated fitness functions and offers these as suggestions to the human designer. In their user study (n=5) participants reported that the suggestions acted as useful sources of inspiration.

\subsection{Methods are Required to Select the Set of Examples to Show the Designer}
As noted by Swearngin et al.~\cite{Swearngin2020-xt} when generating designs using algorithms an essentially unlimited set of examples could be created (within the constraints of the problem). Developers of gallery based human--AI collaborative tools must then decide how to select which examples to present to the designer. Some approaches select examples based on the current state of the design created by the human designer, for example Ngoon et al.~\cite{Ngoon2018-iu} select example feedback statements based on the semantic aspects of the current feedback from the user, and in GUIComp~\cite{Lee2020-db} a number of examples are selected which are measured as similar to the designers current design. Other approaches simply generate a large number of designs and present those which are deemed highest quality, such as in Scout~\cite{Swearngin2020-xt}.

Increasingly the video games research community has turned towards using quality-diversity algorithms, such as MAP--Elites \cite{Mouret2015-kz}, to create a diverse set of high quality examples \cite{Pugh2015-ra} to present to the designer. Quality-diversity (QD) algorithms are a subset of evolutionary algorithms which aim to create a diverse set of high quality solutions to a problem \cite{Pugh2015-ra}. Modern QD algorithms have roots in methods developed to solve multi-modal function optimisation problems, such as niching \cite{Mahfoud1996-qn}, where it is important for the algorithms to maintain the quality and location of multiple solutions throughout the domain. One of the most common QD algorithms, which has spawned numerous variants, is MAP--Elites \cite{Mouret2015-kz}. Algorithms based on Multi-dimensional Archive of Phenotypic Elites (MAP--Elites) create a map of high-quality solutions throughout a space whose dimensions are user defined and may not directly correspond to inputs and outputs of the function being optimised. Crucially, for the context of creating algorithms to support designers, MAP--Elites allows the designer to specify a meaningful `design possibility' space to map. Once the dimensions of interest are defined, a number of bins are generated for each dimension and the best design found so far for each bin is stored in an archive. The current set of best designs for each dimension can then be used as a set of example designs to present to the human designer.

MAP--Elite algorithms have been increasingly applied to gallery based mixed-initiative systems. An early example of this is presented by Alvarez et al. \cite{Alvarez2019-oc}, in an approach they call interactive constrained MAP--Elites (IC MAP--Elites) applied to dungeon map design. IC MAP--Elites allows the designer to select from a set of predetermined dimensions to use for the MAP--Elite algorithm. The underlying evolutionary algorithm then runs for a set number of generations and presents the elites to the designer as suggestions. MAP--Elites have also been applied to the design and balancing of the card game Hearthstone \cite{Fontaine2019-fy} illustrating their potential wide application to many aspects of design and creativity. The MAP--Elite methodology can also be used to assist in the exploration and expansion of crowd sourced content. For example, in Baba is Y'all users can submit level designs which are then allocated to a cell in the MAP--Elites matrix \cite{Charity2022-bl}. Users are then guided towards designing levels which would fill gaps in the MAP--Elites matrix.

\subsection{Our Understanding of how the Method of Selection Affects the Human Experience is Limited}
%Video games research focusses on algorithm
There has been increasing interest in developing and performing user studies to evaluate the effect of mixed-initiative systems~\cite{Alvarez2021-wr, Walton2021-vi, Chan2022-gh}. Much of the research within the video games community focuses on evaluating the performance of the algorithm without much consideration of the human designer experience. For example, Alvarez et al. \cite{Alvarez2021-wr} focus on evaluating the effect the human has on the stability of the MAP--Elites algorithm used to select example designs. In contrast to the video games community's focus, much of the research in HCI focuses on evaluating the human experience using a single predetermined method to select example designs to populate a gallery, such as in \cite{Lee2020-db, Swearngin2020-xt}. Both communities leave a gap of understanding in the effect of the method used to populate the gallery on the human experience~\cite{Craveirinha2015-hm, Chan2022-gh} which our work in this paper aims to explore.

\section{Methodology}
To better understand the effect of gallery based mixed-initiative tools on the creative process we created and evaluated \emph{The Genetic Car Designer}. A common challenge faced by researchers investigating human behaviour is participant recruitment. Secretan et al. \cite{Secretan2011-xq} solved the challenge of recruitment by designing a system which was interesting to, and understandable by, the general public and building a community of self-motivated participants. We adopted a similar approach and designed a gallery based mixed-initiative design tool based on a popular web toy HTML5 Genetic Cars \cite{Matsunaga_undated-su}, where users observe an evolutionary algorithm design and simulate simple cars in real time.

\subsection{The Genetic Car Designer}
\begin{figure}
\centering
\includegraphics[width=.7\textwidth]{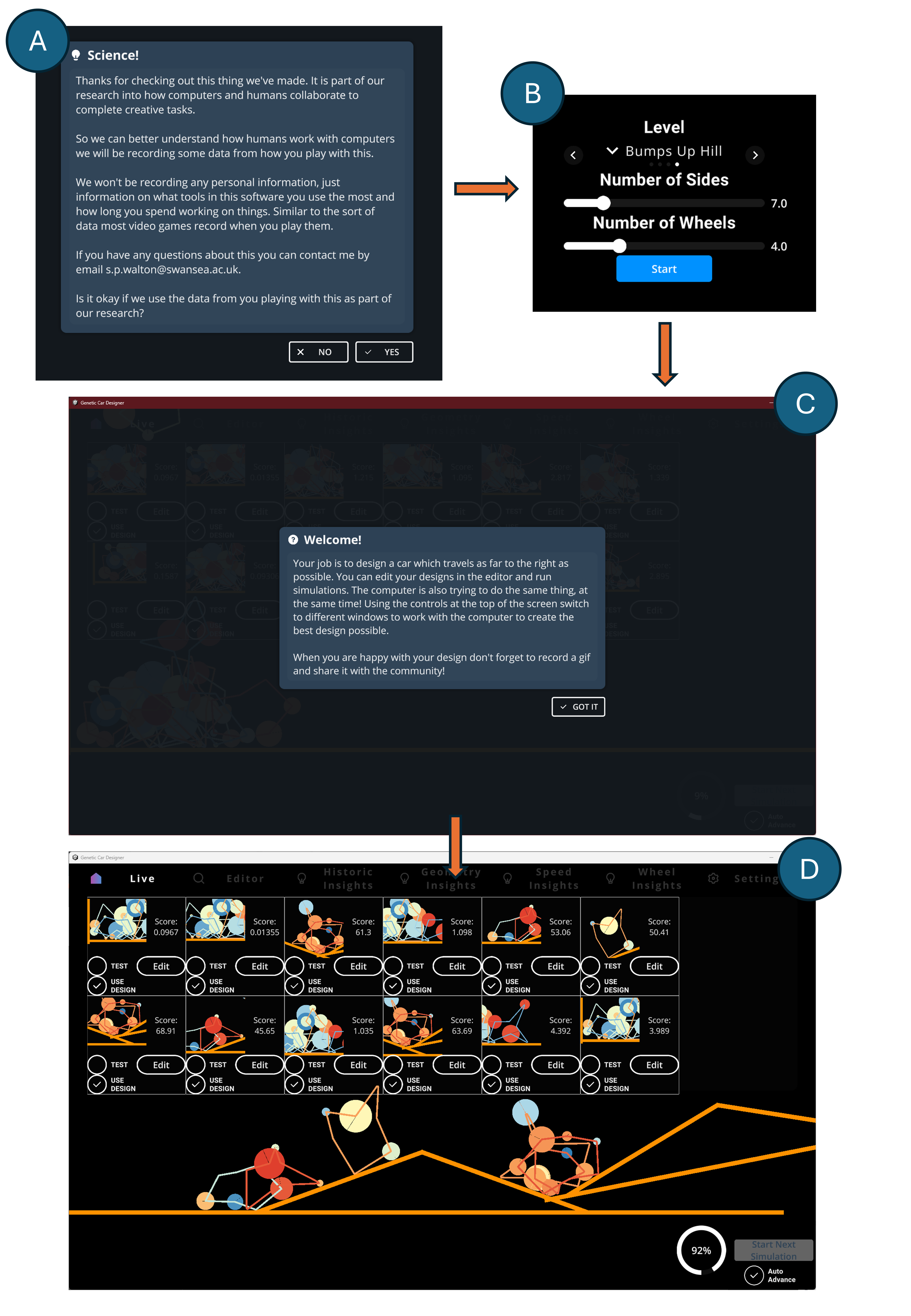}
\caption{The user journey when first launching the genetic car designer. Firstly, the user is asked if they consent to their data being part of the research (A), then the user selects the course or level they wish design a car for along with the number of design dimension they will have (B), they are then given a design brief (C) before being presented with the live view (D). In the main view there is a toolbar at the top of the screen to allow the user to navigate through alternate views.}
\label{fig_onboard}
\end{figure}

\emph{The Genetic Car Designer}, available online\footnote{https://pillbuginteractive.itch.io/genetic-car-designer}, is a mixed-initiative tool for designing a simple 2D approximation of a car, based on a gallery of designs approach. Users can select the course for which they will be optimising the car's performance, along with the number of design variables which define the car (see Figures~\ref{fig_onboard}b and~\ref{fig_car_course} and Section~\ref{sec_designTask}). Once the design objective is explained (see Figure~\ref{fig_onboard}c) the user is presented with a live gallery view (see Figure~\ref{fig_onboard}d and Section~\ref{sec_live}) of the current generation of designs being simulated and evaluated simultaneously as part of the evolutionary algorithm explained in Section~\ref{sec_GA}. In the background of all views is the live simulation of the current generation of designs. The camera locks onto the design which is currently leading. In the bottom right is a percentage indicator showing the progress of the current simulation, and controls for starting the next simulation. By default the simulation is set to auto advance meaning that as soon as one simulation ends the next begins. The user can then select one of the specific views using the interface at the top of the screen. The available views are discussed in detail below, offering functionality to create and edit designs (see Section~\ref{sec_editor}) and examine suggestions from the AI in various gallery views (see Section~\ref{sec_gallery}).

\subsubsection{The Design Task} \label{sec_designTask}
\begin{figure}
%\captionsetup[subfigure]{font=scriptsize}
    \centering
    \begin{subfigure}[t]{0.24\textwidth}
    \centering
    \includegraphics[width=0.95\linewidth]{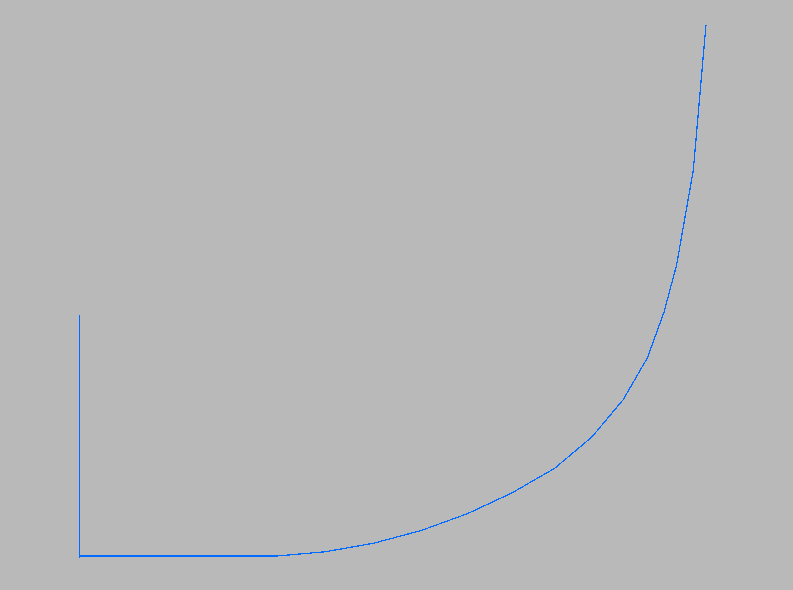}%
    \caption{The hill climb course.}
    \label{fig_hill_climb}
    \end{subfigure}
    \begin{subfigure}[t]{0.24\textwidth}
    \centering
    \includegraphics[width=0.95\linewidth]{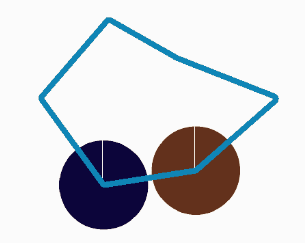}%
    \caption{A car with $6$ vertices and $2$ wheels.}
    \label{fig_MAP_single}
    \end{subfigure}
    \caption{An example of one of the courses participants can select from when starting the task. Cars are dropped into the course at the far left and simulated for 30 seconds. The quality of a design is then measured by the signed distance travelled along the horizontal axis from the designs initial contact point with the ground and the final resting point. The colours of each wheel and the car's body is mapped to the mass of each component. Figure colours have been inverted for clarity.}
    \label{fig_car_course}
\end{figure}

Users are given the design task (see Figure~\ref{fig_onboard}c) which is to design a simple car to travel as far as possible, within a fixed length of time, on a selection of courses. To encourage users to spend longer periods of time with the artefact, we provided four different `courses' to design vehicles for an example of which is shown in Figure~\ref{fig_hill_climb}.

An example of a car design is shown in Figure~\ref{fig_MAP_single}. Each car body is defined by a closed polygon with $N_{v}$ vertices, for a given design task $N_{v}$ is fixed and user specified, where $3 \leq N_{v} \leq 24$. Each vertex $i \in 1...N_{v}$ is positioned at the polar coordinates $(r_{i}, \phi_{i})$ where: $\phi_{i} = 360{(1-i)}/{N_{v}}$ in degrees and each radius, $r_{i}$, a degree of freedom. A closed collision mesh is generated using the vertices which, depending on the relative values of $r_{i}$, may be convex or non-convex. This mesh will collide with the course geometry but not the wheels of the car. A centre of mass is calculated for each car depending on its shape which, along with the car body mass, $M_{body}$, will affect the dynamics of the vehicle. $M_{body}$ is a degree of freedom. Each car has $N_{w}$ wheels which is fixed and user specified for a given design task, where $1 \leq N_{w} \leq 12$. Wheels are implemented using the wheel joint 2D component in the Unity3D game engine\footnote{https://docs.unity3d.com/Manual/class-WheelJoint2D.html} which simulates a motor powered wheel connected to a body via a suspension spring. Every wheel has $5$ degrees of freedom associated with it:
\begin{enumerate}
    \item The location of the wheel. This is defined by the vertex $i$ to which the wheel is attached. More than one wheel may be attached to the same vertex.
    \item The radius of the wheel.
    \item The mass of the wheel.
    \item The target speed of the motor attached to this wheel. A torque is applied dynamically during the simulation in an attempt to reach this speed.
    \item The oscillation frequency of the suspension attached to this wheel, which controls the stiffness of the spring.
\end{enumerate}
Participants in the large-scale study were given the option to change $N_{v}$ and $N_{w}$ before attempting the design task. The total number of degrees of freedom (or dimensions in the design problem), $D$, can be calculated using: $D = 1 + N_{v} + 5N_{w}$. Each degree of freedom is constrained within defined bounds selected to ensure the designs created are physically valid.

Designs were tested numerically using the real-time two-dimensional rigid body physics simulation in Unity3D\footnote{https://unity.com/}. To evaluate a design it is dropped onto the course and released to freely travel based on the physics simulation. Cars are dropped from a height above the course and the quality metric, or objective function, is the signed distance, along the horizontal axis, from the first point the car hits the course to the final position at the end of a 30-second run. Using the signed distance ensures that cars which move in the wrong direction have a negative quality. The quality metric is given the term "Score" when presented to users in the gallery views.  

\subsubsection{Mixed-Initiative Evolutionary Algorithm} \label{sec_GA}
\begin{figure}
\centering
\includegraphics[width=.7\textwidth]{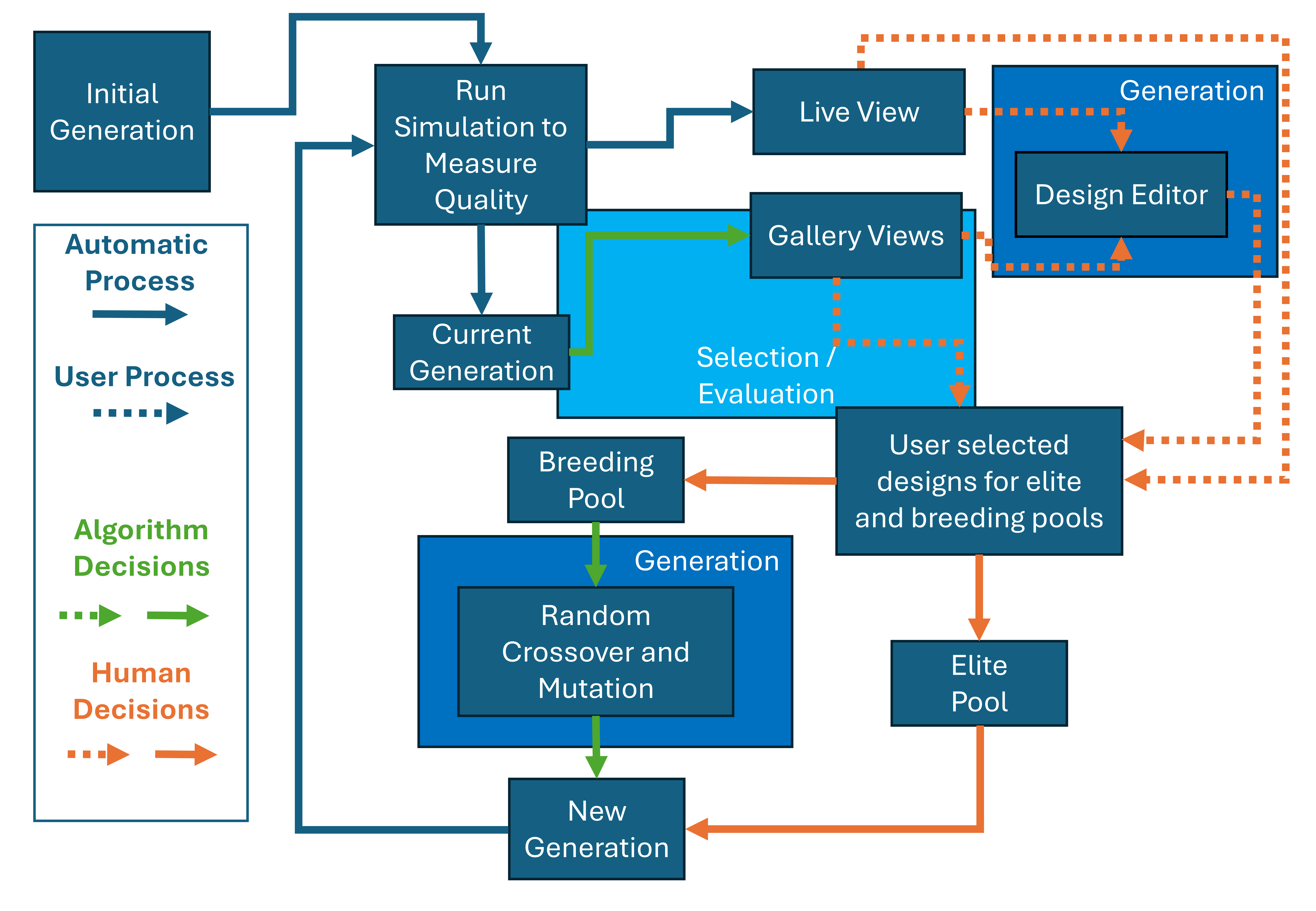}
\caption{A flow diagram illustrating the optimisation algorithm and the decision making roles of the human and the algorithm. Here the algorithmic decision of which designs to populate the galleries with is highlighted since it is the core focus of our study.}
\label{fig_GA_flow}
\end{figure}
Figure~\ref{fig_GA_flow} presents an overview of the underling algorithms which drive the Genetic Car Designer. A central concept in the system is a generation of designs, each design defined by the degrees of freedom and quality metric defined in Section~\ref{sec_designTask}. The system runs a simulation of the current generation in parallel to measure the quality of each design, then through a combination of human and algorithm decisions a new generation is generated ready for simulation.  The process iterates for as long as the user wishes. In our implementation the number of designs in a generation is kept as a constant $12$, this value was selected to ensure that the live and gallery views did not take up too much of the screen space (see Figure~\ref{fig_onboard}d). The initial generation is generated by randomly initialising each degree of freedom for $12$ designs as described in Section~\ref{sec_designTask}. 

To generate the next generation after the current generation's simulation is complete the optimiser uses designs from the elite and breeding pools of designs. The pools are first emptied and then the optimiser requests designs from the design editor, live view and all gallery views which have been marked by the user as "Test" or "Use Design". In Figure~\ref{fig_live_view_single} the two check boxes the user can use to mark designs are shown. When mousing over these check boxes a tool tip describes the functionality as "Tell the computer to test this design in the next simulation" and "Tell the computer to use this design to come up with new ideas" for "Test" and "Use Design" respectively. When the user first launches the application all individuals in the Live View are marked as "Use Design", but no other check boxes are checked.

Designs marked as "Test" are entered into the elite pool, these designs are entered into the new generation without any modification. The optimiser then generates new designs using the breeding pool, which is made up of designs marked as "Use Design", until the next generation has $12$ designs. Two mutually exclusive pairs of designs are selected from the breeding pool and the two best quality designs from these pairs are selected as parents. In the case there are fewer than four designs in the breeding pool the algorithm adds random designs from the last generation until there are four. A new child design is generated through crossover and mutation operations performed on the two selected parents. Firstly, for each degree of freedom the child is given a uniformly random value between the parents' degrees of freedom. Secondly, there is a 10\% mutation chance of a degree of freedom to be randomly initialised. Once $12$ designs have been added to the next generation, the next simulation starts.

\subsubsection{The Live Views} \label{sec_live}
\begin{figure}
%\captionsetup[subfigure]{font=scriptsize}
    \centering
    \begin{subfigure}[t]{0.7\textwidth}
    \includegraphics[width=.9\linewidth]{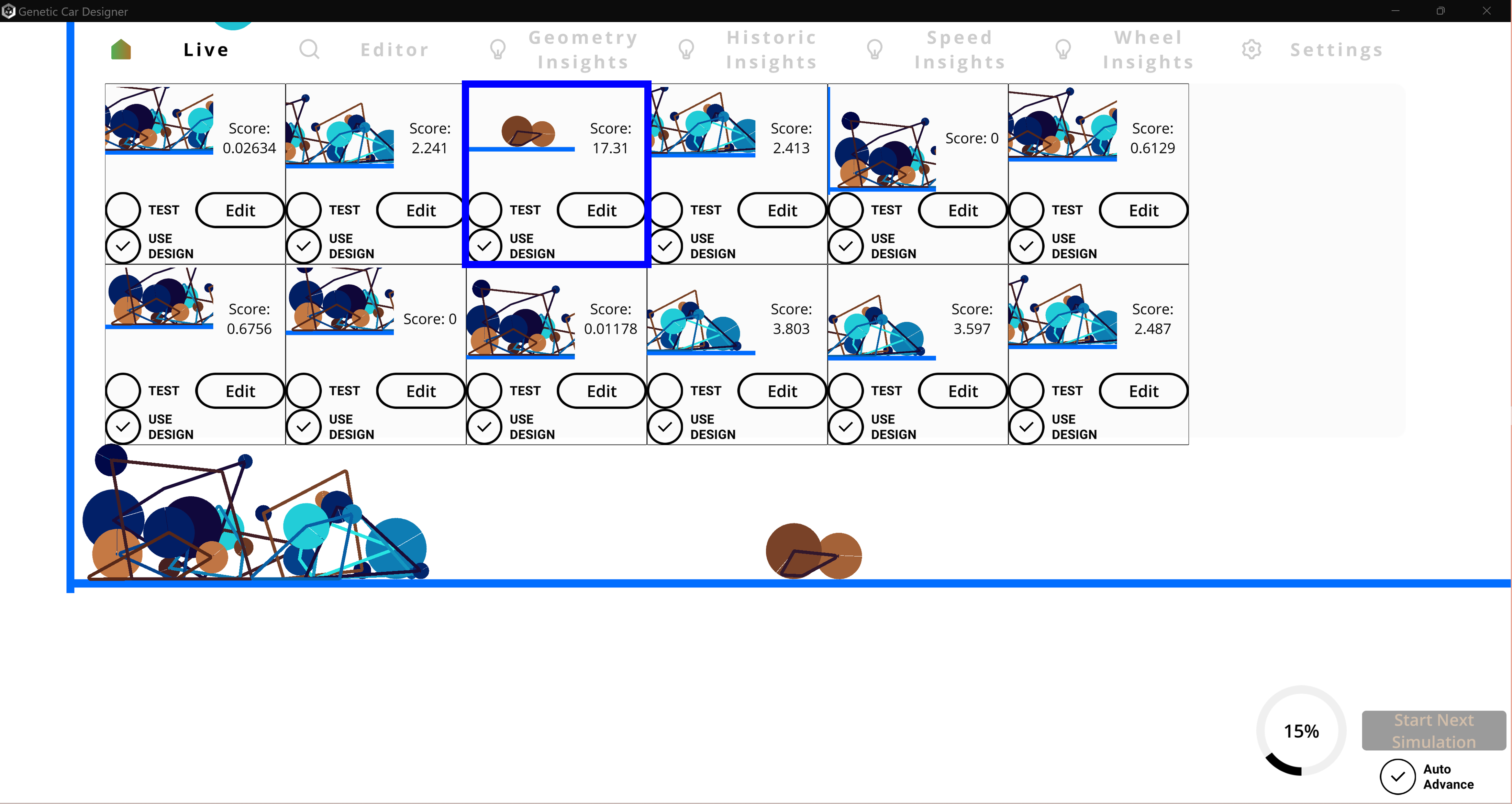}%
    \caption{The Live view which was the default view for participants.}
    \label{fig_live_view} % label needs to come after caption in figure*
    \end{subfigure}
    \begin{subfigure}[t]{0.3\textwidth}
    \centering
    \includegraphics[width=\linewidth]{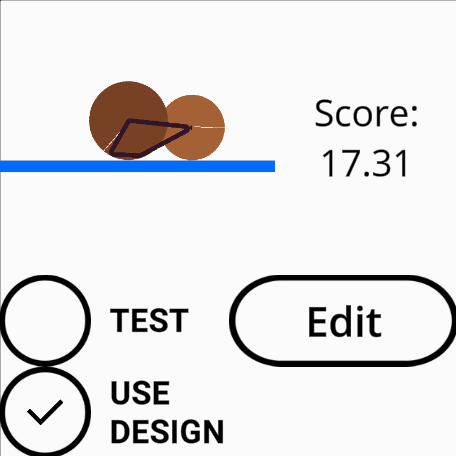}%
    \caption{Zoom of a single design in live view.}
    \label{fig_live_view_single}
    \end{subfigure}
    \caption{The default live view which shows each design in the current generation as they are being evaluated. From here the user can select which designs to use to create the next generation and which to test in the next generation. They can also click edit to edit any design in the edit view. This is a real time camera showing the actual position of the design in the current simulation. Figure colours have been inverted for clarity.}
\end{figure}
The first view presented to the user is the Live view which is shown in Figure~\ref{fig_live_view}. It is made up of $12$ individual views (see Figure~\ref{fig_live_view_single}) showing the current live position of each design in the current generation along with its current score (the fitness value).

\subsubsection{The Gallery Views}\label{sec_gallery}
\begin{figure}
\centering
\includegraphics[width=.9\textwidth]{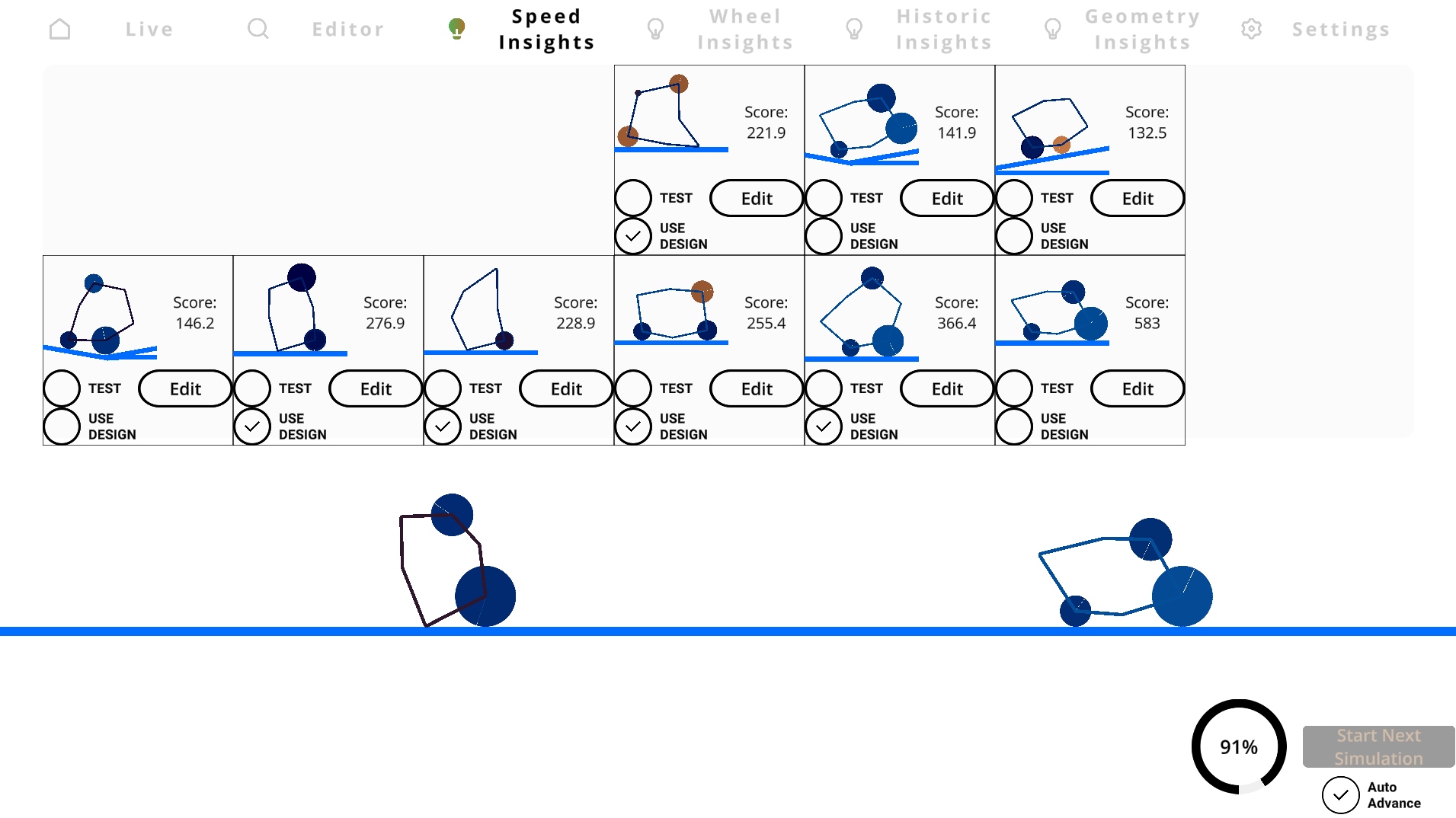}
\caption{An example of a Gallery View. This is showing the Speed Insights, note that there are three blank gallery thumbnails which indicates part of the search space where designs have not been tested. Unlike the live view, the thumbnails here are static images from the last time step in the simulation where these designs were tested. Figure colours have been inverted for clarity.}
\label{fig_gallery}
\end{figure}
The Gallery Views are presented to the human designer in the same style as the live view, the key difference being that the designs presented are not currently being simulated, and a snapshot of the design at the end of a simulation is presented. The same controls are given for each design in a gallery as in the individual Live views, but by default no check box is selected. 

For the MAP--Elites based galleries a metric is defined and $12$ bins between the minimum and maximum values for that metric are defined. The minimum and maximum values for each metric are initially estimated, then updated if designs are found which fall outside these values. There are three galleries which are controlled by the MAP--Elites algorithm: \emph{speed insights}, \emph{wheel insights} and \emph{geometry insights}. The metrics, which were not revealed to participants, for these MAP--Elite galleries are as follows:
\begin{itemize}
    \item Speed insights: The mean speed of the wheels on the car.
    \item Wheel insights: The mean radius of the wheels on the car.
    \item Geometry insights: The mean signed distance (along the position vector of the vertex) from each vertex to the centre of mass (centre of mass calculations include the wheel masses) 
\end{itemize}
Each generation the optimiser replaces a design in a MAP--Elites bin if a design of better quality in that bin has been found. In summary, MAP--Elites organises the design possibility space into a grid of diverse and high performing designs. Presenting a wide range of design possibilities to users in this structure affords exploration by supporting designers to overcome fixation on a narrow area of the design space.

In addition to the views which use MAP--Elites to generate suggestions, the tool also included a view labelled \emph{historic insights} which acts as our experimental control. Designs presented in the \emph{historic insights} are designs selected at random from all designs tested. The order in which these views are presented to participants in the navigation bar, which can be seen at the top of Figure~\ref{fig_live_view}, is randomised each time the software is launched.

\subsubsection{The Editor View} \label{sec_editor}
\begin{figure}
\centering
\includegraphics[width=.9\textwidth]{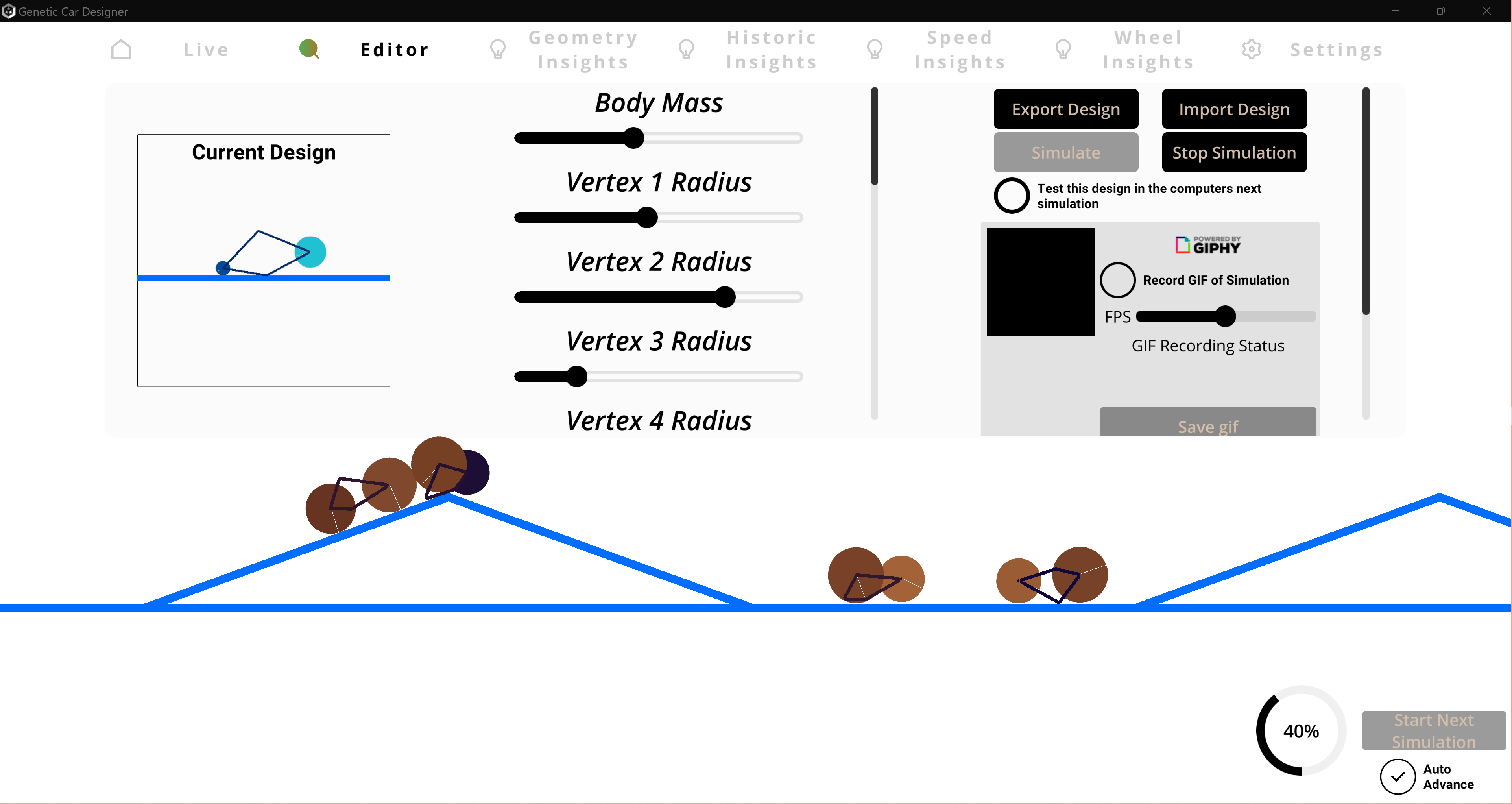}
\caption{The editor where a user can modify existing designs and create their own. A list of sliders allows the user to change each gene, with the image on the left updating in real time. The user can then export the design, add it to the next generation or simulate it without adding it to the generation. Figure colours have been inverted for clarity.}
\label{fig_edit}
\end{figure}
The editor view, shown in Figure~\ref{fig_edit} allows the human designer to edit existing and create new designs. In the Live and Gallery views a user can click the edit button on an individual design to open it in the editor, but the user can also simply select the editor from the toolbar at the top of the screen. Each degree of freedom is edited using a list of sliders, when the user mouses over the slider handle the numerical value of the degree of freedom is displayed. The current design updates as the user interacts with each slider and at any point the user can click a button to run a simulation of the design. During a simulation if the user edits a degree of freedom the simulation stops. When happy with the design the user can include the design in the next generation of the genetic algorithm by checking the labelled box.

\subsection{Experimental Approach}
To understand the benefits and limitations of gallery based mixed-initiative design tools, and to examine what impact (if any) the algorithm used to generate the gallery of examples has on the human experience, we conducted two within-subjects studies: (1) a large-scale quantitative field study and (2) a small-scale mixed-methods lab study. Our studies were designed to answer the following research questions:
\begin{itemize}
    \item RQ1: Do galleries of examples influence user engagement with the design process?
    \item RQ2: Do galleries of examples have an influence on the quality of designs produced?
    \item RQ3: Do galleries of examples generated using MAP--Elites have a different utility compared to a random set of suggestions?
\end{itemize}
\subsubsection{Defining Engagement}
To answer RQ1 we need to define what we mean by user engagement, which is less straight forward than defining quality of designs (RQ2) and the comparative utility of the galleries (RQ3). For the purposes of this study engagement is defined as a variable state of the human designer which can be categorised as Cogitative, Emotional or Behavioural. Although Sidner et al.~\cite{Sidner2005-nu} define engagement as a \emph{process} by which two or more entities maintain a connection, it is more practical to consider it as a variable state~\cite{Doherty2019-ee} characterised by of a responsive dyadic form of interaction. To operationalise the concept of engagement, and identify suitable proxies/measures, Doherty et al.~\cite{Doherty2019-ee} suggest asking if the engagement of interest is Cognitive, Emotional and/or Behavioural. Cognitive engagement is focussed on conscious aspects such as attention, awareness and effort. Emotional engagement is focussed on the subjective experience, such as expressions of boredom, interest and feelings. Behavioural engagement emphasises action and participation, which can also include physical behaviour. Throughout the presentation of the results we will identify proxies for the different aspects of engagement defined above.

\section{Field Study}
Our primary aim with the field study was to collect a large data set to help answer our research questions. To meet our target of a large sample size the conditions were less controlled than a typical user study. Ethical approval was given to carry out the field study by the Swansea University Faculty of Science and Engineering Ethics Committee (SU-Ethics-Staff-030822/505).

\subsection{Procedure}
An unmodified version of The Genetic Car Designer was made freely available to the general public on itch.io\footnote{https://itch.io}, an online marketplace for hosting a wide range of video games and related artwork. Recruitment was carried out via several social networks, with the majority of participants coming from Reddit. Participants were asked if they were willing to take part in the study when launching the tool, if they agreed then analytics data from each design session was uploaded to our server for analysis; otherwise no data was recorded.

Upon launching the application participants were given the design task shown in Figure~\ref{fig_onboard}c. Participants were free to spend as long as they wished on the task and could perform the task multiple times with different courses and different numbers of degrees of freedom. Data uploaded for analysis was done on a per session basis, and to ensure anonymity and compliance with data protection laws we are unable to detect a single participant uploading data from multiple sessions.

To compare groups of data we used a Mann--Whitney U test, rejecting the null hypothesis that there is no difference between groups if $p<\alpha_{adjusted}$ with adjusted significance level $\alpha_{adjusted}= 0.002$. This significance level was calculated using the Bonferroni correction for multiple comparisons with $\alpha_{initial}=0.01$ and five comparisons.
To evaluate correlation we use a Spearman rank-order correlation coefficient, rejecting the null hypothesis that there is no correlation if $p<0.002$. Data distributions are presented below using kernel density estimate plots, which show the distribution of data by plotting the probability density with a set smoothing kernel \cite{Vermeesch_2012}.

\subsection{Results}
%Should there be a small paragraph here?
\subsubsection{Participants}
Data for the large scale study was collected within a predetermined time period between 4th August and 4th September 2022. In total the tool was launched 2,493 times during the data collection period. Only data from participants who consented to take part in the study was uploaded to our server to make the initial data set. Data was then removed from this set for one of the following reasons:
\begin{itemize}
    \item The version of the tool used was not the most recent version. Prior to 4th August a number of bugs were identified with the software and subsequently fixed.
    \item An upper bound on the maximum distance a car could travel was calculated. Any data point which had a final fitness greater than this value was removed. Unrealistic fitness values could be caused by divergence and inaccuracies in the physics model, or by users who modified the tool to create false values. Since this study was not conducted in lab conditions it is possible that users could reverse engineer the software and post modified data.
    \item Data entries which contained NaNs were also removed, this indicates errors in the physics engine as discussed above.
\end{itemize}
Once invalid data was removed we were left with a data set of 808 sessions. In 50\% of these sessions the participants were simply passive and let the evolutionary algorithm run without offering any suggestions or input to the algorithm, showing zero behavioural engagement (i.e. no action or participation). In 36\% of the 808 sessions, participants only offered feedback to the algorithm by suggesting designs they created in the editor, and in 14\% of the sessions participants offered feedback to the algorithm using both the editor and gallery views. In the 404 non-passive participant sessions, 54\% of cases participants only opened the editor view and 46\% of cases participants viewed at least one gallery in addition to the editor. 

\paragraph{Passive participants selected more complex design problems.} As detailed in Section~\ref{sec_designTask} participants could select the number of degrees of freedom, or dimensions, in the design task. The mean number of dimensions was 25 distributed between the minimum of 9 and maximum of 85. There was no statistical difference between the number of dimensions selected by those who interacted with only the editor and those who interacted with at least one gallery view. However, participants who did not engage behaviourally with the tool on average selected more degrees of freedom when starting the task ($p<0.002$) than those who did.

\subsubsection{Engagement}
\begin{figure}
\centering
\includegraphics[width=.5\linewidth]{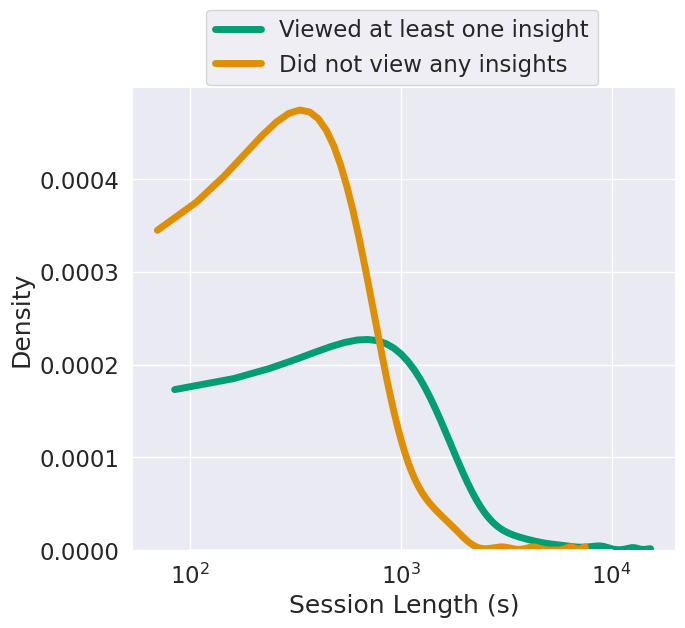}%
\caption{A kernel density estimate plot showing the distributions of session length for each group of participants based on viewing behaviour.}
\label{fig_sessionlength_viewTime}
\end{figure}

\begin{figure}
    \begin{subfigure}[t]{0.49\textwidth}
    \centering
    \includegraphics[height=8.5cm]{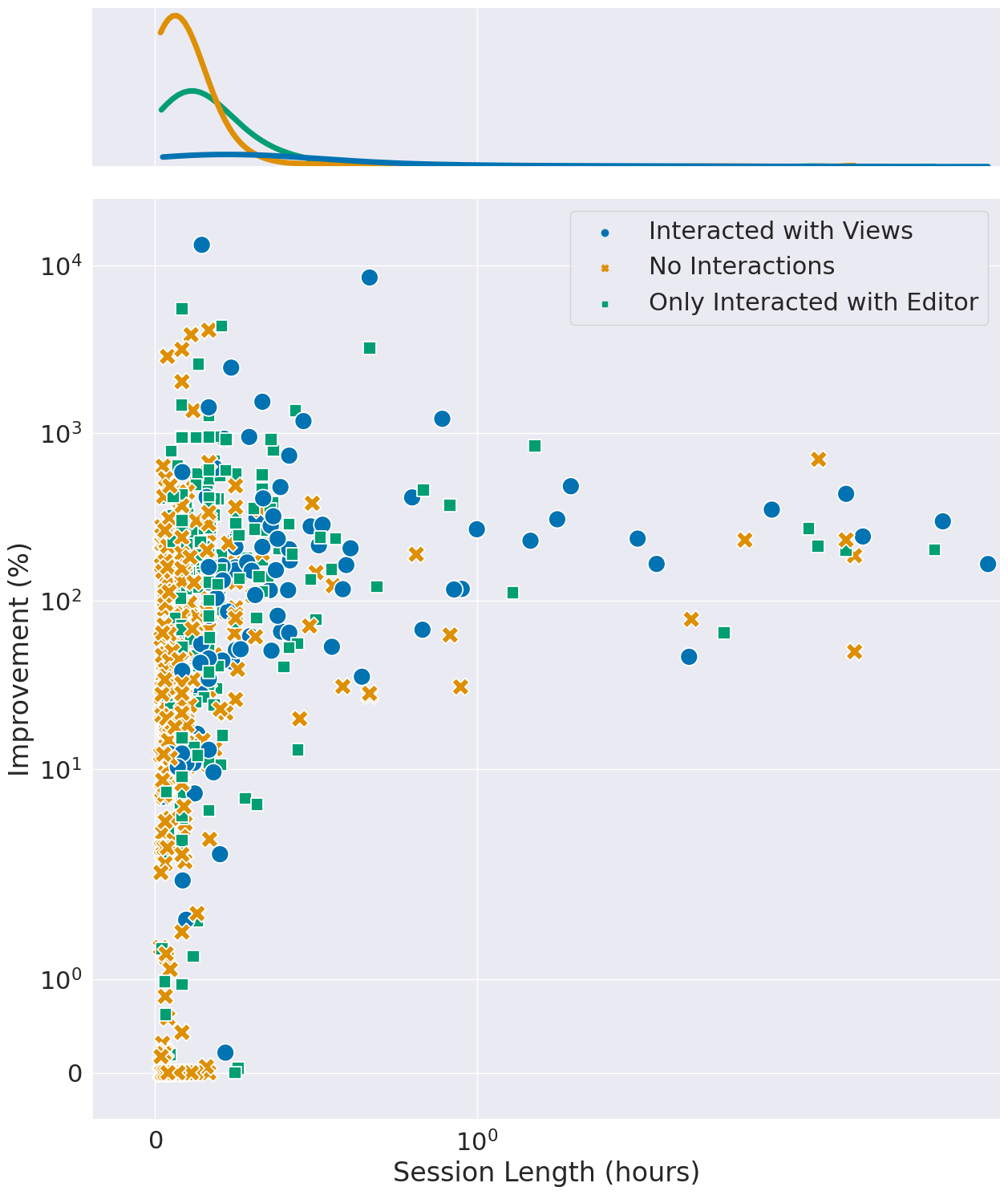}%
    % \caption{Scatter plot showing the session length on the x-axis and percentage improvement in that design session on the y-axis. The symbol of each point indicates which group the participant fell into.}
    \caption{A scatter plot showing participants' session time against their improvement.}
    \label{fig_timevsimprovement}
    \end{subfigure}
    \begin{subfigure}[t]{0.49\textwidth}
    \centering
    \includegraphics[height=8.5cm]{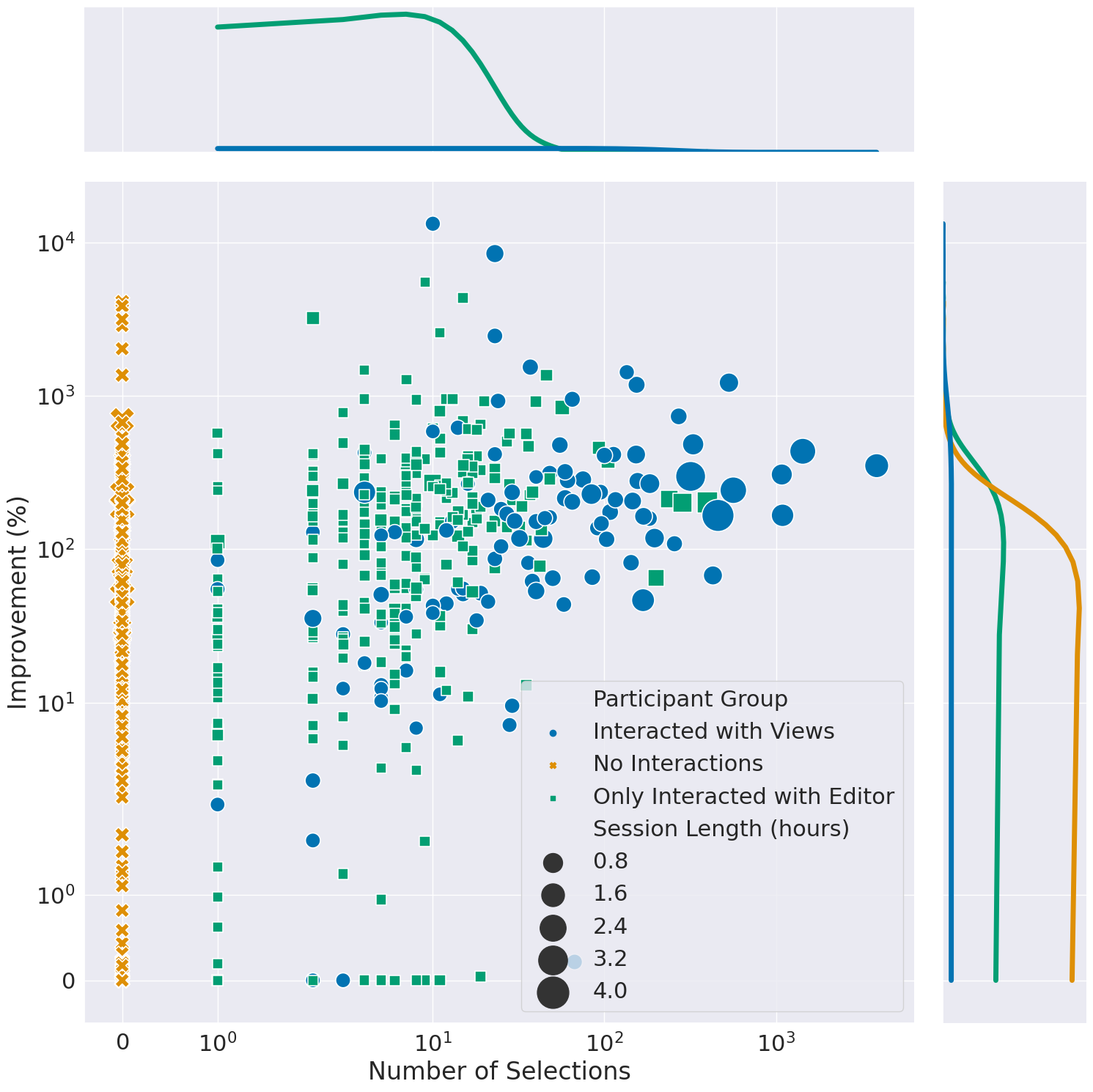}%
    % \caption{Scatter plot showing the number of selections in a session on the x-axis and percentage improvement in that design session on the y-axis. The symbol of each point indicates which group the participant fell into, and the size of the point indicates the amount of time spent in this session.}
    \caption{A scatter plot showing participants' number of selections made against their improvement. The size of the data indicates corresponds to session length.}
    \label{fig_selectsvsimprovement}
    \end{subfigure}
    \caption{Scatter plots of session length and number of selections compared to total improvement. Respective kernel density estimate plots are shown above with the performance improvement kernel density estimate plot shown on the right.}
\end{figure}

Since the design task had no fixed end point and a participant could exit at any time, an assumption is made that the length of time each session lasted can be used as a measure of attention and by proxy cognitive engagement. Sessions lasted between 1 minute and 4.3 hours, with a mean time of 12 minutes. The full distribution for each group of participants is shown in Figures ~\ref{fig_sessionlength_viewTime} and ~\ref{fig_timevsimprovement}.

\paragraph{Participants who collaborated with the algorithm were more cognitively engaged in the task.} Passive participants were the least cognitively engaged with the task, followed by those who just interacted with the editor, and finally the participants who interacted with the editor and gallery views were the most cognitively engaged in the task ($p<0.002$ for all comparisons). If we consider this collaboration with the algorithm as a proxy for behavioural engagement, then these results show a connection between behavioural and cognitive engagement.

\paragraph{Participants who viewed the galleries were more cognitively engaged in the task than those who did not.} When comparing participants based on their view behaviour, participants who viewed at least one gallery spent significantly ($p<0.002$) longer cognitively engaged with the task than others. 

\begin{table}
% increase table row spacing, adjust to taste
% if using array.sty, it might be a good idea to tweak the value of
% \extrarowheight as needed to properly center the text within the cells
\caption{Mean session length by participant group. The error presented is the standard error in the mean.}
\label{table_length}
\centering
\begin{tabularx}{\linewidth}{X X}
\toprule
Group & Session Length (s) \\
\midrule
No interactions & $439\pm50$ \\
Only interacted with Editor & $728\pm71$ \\
Interacted with at least one gallery & $1652\pm222$ \\
\midrule
Did not view any gallery & $562\pm60$ \\
Viewed at least one gallery & $1351\pm140$ \\
\bottomrule
\end{tabularx}
\end{table}

\subsubsection{Design Quality and Effectiveness of Design Sessions}
We wanted to understand if providing galleries of examples would influence the quality of the designs produced. The quality, or fitness, of a design is the signed distance travelled along the horizontal axis---this raw value is hard to compare between two sessions, however, due to the random initialisation. Instead, to measure the effectiveness of a particular design session, the improvement from the best design in the initial randomly generated set of designs to the best design found during the session was calculated as a percentage. The mean improvement from all sessions was 208\% and improvements ranged from 0\% to 13,241\%. 

\paragraph{The higher the initial design quality, the harder it was to make an improvement.} The fitness of the best performing design in the first generation of the evolutionary algorithm was recorded for each session. Initial finesses ranged from $2.95$ to $467$ with a mean of $142$. We found no relationship between the initial design quality and participant behaviour. There is, however, a negative correlation ($p<0.002$) between initial design quality and overall improvement, the correlation coefficients are presented per group in Table~\ref{table_initialFitness}, using Fishers z-transformation no statistical difference between these correlations is found. This difference in starting design introduces a potential confound into the field study, which we eliminate from the lab study by introducing a fixed set of starting designs. When investigating the distributions of initial design quality we found no statistically significant difference between the different groups of participants based on their behavioural engagement. The lack of statistical difference in initial design quality between groups gives us confidence that comparisons of percentage improvements are not significantly affected by the confound identified. 
\begin{table}
% increase table row spacing, adjust to taste
% if using array.sty, it might be a good idea to tweak the value of
% \extrarowheight as needed to properly center the text within the cells
\caption{Spearman rank-order correlation coefficients for fitness of the best performing design in the first generation vs percentage improvement. Negative correlation implies that overall improvement (as a percentage) decreases as the quality of the initial design increases. p<0.002 in all cases.}
\label{table_initialFitness}
\centering
\begin{tabularx}{\linewidth}{X X}
\toprule
Group & Correlation Coefficient\\
\midrule
No interactions & $-0.616$ \\
Only interacted with Editor & $-0.599$ \\
Interacted with at least one gallery & $-0.504$ \\
\midrule
Did not view any gallery & $-0.575$ \\
Viewed at least one gallery & $-0.565$ \\
\bottomrule
\end{tabularx}
\end{table}

\paragraph{Longer design sessions were more effective.} Figure~\ref{fig_timevsimprovement} shows the percentage improvement in design plotted against session length. There is a positive correlation between session length and improvement ($p<0.002$) for all groups of participant. 

\begin{table}
% increase table row spacing, adjust to taste
% if using array.sty, it might be a good idea to tweak the value of
% \extrarowheight as needed to properly center the text within the cells
\caption{Mean percentage improvement in design achieved in each session separated by participant group. The error presented is the standard error in the mean.}
\label{table_improvement}
\centering
\begin{tabularx}{\linewidth}{X X}
\toprule
Group & Mean Improvement (\%) \\
\midrule
No interactions & $124\pm19$ \\
Only interacted with Editor & $243\pm30$ \\
Interacted with at least one gallery& $420\pm140$ \\
\midrule
Did not view any gallery & $197\pm35$ \\
Viewed at least one gallery & $373\pm77$ \\
\bottomrule
\end{tabularx}

\end{table}
\paragraph{Sessions where participants collaborated with the design tool were more effective than the tool working independently.} Table~\ref{table_improvement} shows the mean improvement in design quality for each group of participant. Sessions where participants engaged behaviourally with either the editor and/or a gallery had significantly ($p<0.002$) higher improvements of quality than those where participants were passive and let the evolutionary algorithm drive the process.

\paragraph{Sessions where participants viewed at least one gallery were more effective than those where participants just worked with the editor.} When considering sessions with active participants, we found that participants who viewed at least one gallery performed significantly better than those who did not ($p<0.002$).

\paragraph{Sessions where more designs were suggested to the algorithm by the designer were more effective.} The tool keeps track of the number and type of design selections made from each view and the editor, i.e. selecting a design to be entered into the elite and/or breeding pools. Participants made between zero and 3,862 selections per session, although this maximum value was an outlier as the mean was 24 selections. This metric indicates the number of actions taken by the human designer and therefore is a proxy for behavioural engagement. Figure~\ref{fig_selectsvsimprovement} shows the full distributions for each set of groups; all distributions are statistically different from each other ($p<0.002$). There is a positive correlation between number of selections and improvement ($p<0.002$), this correlation is stronger for participants who interacted with the editor and galleries (Rank coefficient $=0.567$) compared to those who just interacted with the editor (Rank coefficient $=0.445$) but this difference is not statistically significant.

\subsubsection{Differences Between Galleries}
We wanted to understand if the algorithm used to generate the gallery of suggestions had an effect on the usefulness of the suggestions, or if simply a random set of suggestions was good enough. To assess this we provided participants access to four galleries, three based on MAP--Elites (i.e. the intelligent algorithm) and one a random selection of designs (i.e. the control). The order in which these galleries are presented to the user is randomised, however we were concerned that the names of the gallery may introduce bias. For each session we recorded which gallery the user first viewed and, upon analysing the data, found that no gallery was statistically more likely to be opened first compared with the others. This gives us confidence that no significant bias was introduced.

\begin{figure}
    \centering
    
    \begin{subfigure}[t]{0.49\textwidth}
    \centering
    \includegraphics[width=\textwidth]{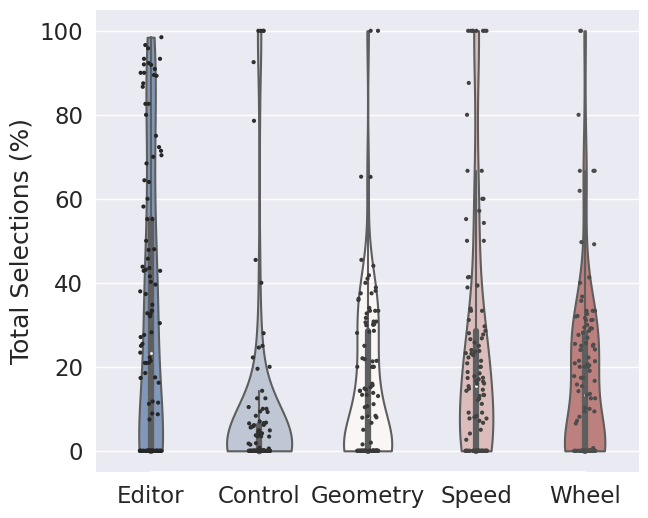}%
    \caption{Violin plot showing the distributions of total number of selections from each view.}
    \label{fig_selectionView}
    \end{subfigure}
    \hfill
    \begin{subfigure}[t]{0.49\textwidth}
    \centering
    \includegraphics[width=\textwidth]{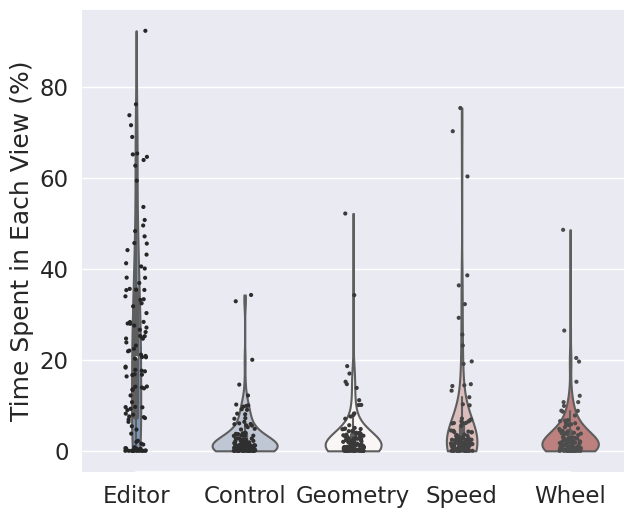}%
    \caption{Violin plot showing the distributions of relative time spent in each view.}
    \label{fig_timeView}
     \end{subfigure}
     \hfill

    \caption{Violin plots indicating in which view participants made their selections and how much time they spent in each view.}
    \label{fig_violin_improve}
     
\end{figure}

\paragraph{Participants spent the same amount of time in MAP--Elite based galleries as the random control, but selected more designs from the MAP--Elite galleries.} The number of selections from and time spent in each view in the tool was recorded. In this part of the analysis we only consider the group of participants who interacted with both the editor and at least one gallery. Table~\ref{tab_views} gives an overview of the data and shows the results of Mann--Whitney U tests comparing each gallery to the control. The distributions are also presented in Figures~\ref{fig_selectionView} and~\ref{fig_timeView}. Most selections were made from the Editor, followed by the Speed, Wheel and Geometry MAP--Elites galleries ($p<0.002$ in all cases when compared to the control). Participants spent the most time in the Editor ($p<0.002$) and there was no observed differences between the time spent in the remaining views. Putting this in the context of engagement, we observe that participants gave equal attention (cognitive engagement) to the MAP--Elite and random galleries but took more actions (behavioural engagement) from the MAP--Elite galleries. 
\begin{table}
 \caption{Overview of data comparing the time spent and selections from each view in the tool. The difference columns refer to the result of a  Mann--Whitney U test. Values in bold are significant ($p<0.002$). The error presented is the standard error in the mean.}\label{tab_views}
    \begin{tabularx}{\linewidth}{X | X X | X X }
    \toprule
    {View} & {Mean Total Selections (\%)} & {Difference to Control} & {Mean Time Spent in View (\%)} & {Difference to Control} \\
    \midrule
    Control & $8.87\pm2.0$ & -- & $3.16\pm0.5$ & -- \\    
    Geometry & \bf{15.0}$\pm$\bf{2.0} & {\bf{Higher }}  & $3.35\pm0.6$ & None  \\
    Speed & \bf{23.2}$\pm$\bf{3.0} & {\bf{Higher }} & $6.31\pm1.0$ & None  \\
    Wheel & \bf{17.3}$\pm$\bf{2.0} & {\bf{Higher }} & $3.59\pm0.5$ & None  \\
    Editor & \bf{32.2}$\pm$\bf{3.0} & {\bf{Higher }} & \bf{59.5}$\pm$\bf{2.0}  & {\bf{Higher }} \\
    \bottomrule
    \end{tabularx}
    
    % \caption{Overview of data comparing the time spent and selections from each view in the tool. The difference columns refer to the result of a  Mann--Whitney U test}\label{tab_views}
    % \begin{tabularx}{\linewidth}{R | C C | C C }
    % \toprule
    % {View} & {Mean Total Selections (\%)} & {Difference to Control} & {Mean Time Spent in View (\%)} & {Difference to Control} \\
    % \midrule    
    % Control & $8.87\pm2.0$ & -- & $3.16\pm0.5$ & -- \\    
    % Geometry & $15.0\pm2.0$ & None ($p>0.0001$) & $3.35\pm0.6$ & None ($p>0.0001$) \\
    % Speed & $23.2\pm3.0$ & {\bf{Higher ($p<0.0001$)}} & $6.31\pm1.0$ & None ($p>0.0001$) \\
    % Wheel & $17.3\pm2.0$ & {\bf{Higher ($p<0.0001$)}} & $3.59\pm0.5$ & None ($p>0.0001$) \\
    % Editor & $32.2\pm3.0$ & {\bf{Higher ($p<0.0001$)}} & $59.5\pm2.0$ & {\bf{Higher ($p<0.0001$)}} \\
    % \bottomrule
    % \end{tabularx}
    % \begin{tablenotes}
    % \item The error presented is the standard error in the mean.
    % \end{tablenotes}
\end{table}
% \begin{figure}[bt!]
% \centering
% \includegraphics[width=.7\linewidth]{selectionsviewperc.png}%
% \caption{Violin plot showing the distributions of total number of selections from each view.}
% \label{fig_selectionView}
% \end{figure}
% \begin{figure}[bt!]
% \centering
% \includegraphics[width=.7\linewidth]{timeperviewperc.png}%
% \caption{Violin plot showing the distributions of relative time spent in each view.}
% \label{fig_timeView}
% \end{figure}

\paragraph{Greater cognitive and behavioural engagement with the galleries resulted in more effective design sessions, and that correlation is stronger for MAP--Elite based galleries than the random control.} Table~\ref{tab_views_correlation} shows correlation coefficients calculated when considering the time spent and selections from each view and performance. The more selections participants made from the wheel, geometry or speed galleries, the more designs the participants suggested from the editor, and the more selections participants made from the all gallery views combined led to increases in performance ($p<0.002$). There was a significant positive correlation between the total proportion of time spent in the non-control gallery views, time spent in the speed and wheel gallery views, and the editor and performance  ($p<0.002$).

\begin{table}
    \caption{Spearman rank-order correlation coefficients for the time spent and selections from each view in the tool vs percentage improvement. Values in bold are significant ($p<0.002$).}
    \label{tab_views_correlation}
    \begin{tabularx}{\linewidth}{X | X | X }
    \toprule
    {View} & {Number of Selections vs Improvement} & {Time Spent in View vs Improvement} \\
    \midrule    
    Control & 0.146 & 0.156  \\    
    Geometry & {\bf{0.246}}  & 0.244  \\
    Speed & {\bf{0.291}}  &  {\bf0.318}  \\
    Wheel & {\bf{0.464 }}& {\bf0.358}  \\
    \midrule
    Control + Geometry + Speed + Wheel & {\bf0.385 }& {\bf0.301}  \\
    \midrule
    Editor & {\bf0.497 }&  {\bf0.353}   \\
    \bottomrule
    \end{tabularx}
\end{table}

\section{Lab Study}
\begin{figure}
\centering
\includegraphics[width=0.9\textwidth]{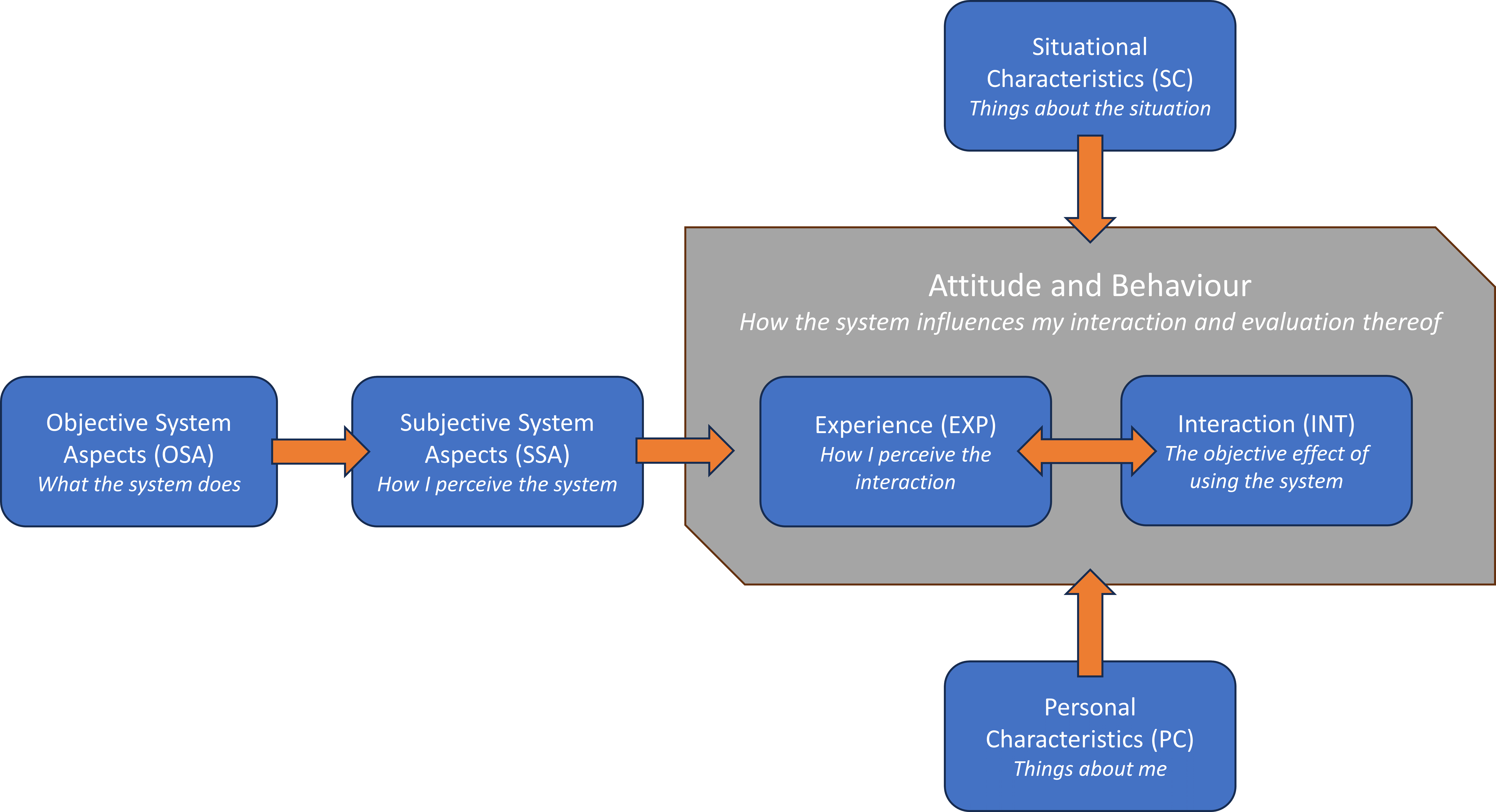}%
\caption{Framework for evaluating recommender systems proposed by Knijnenburg et al. \cite{Knijnenburg2012-hf}. The framework acts as a model of the system, using this context we can pose questions to participants and evaluate their responses.}
\label{fig_framework}
\end{figure}

To enable an in-depth investigation into the trends observed during the large scale field study a small scale lab study was designed. The lab study procedure was designed following the data analysis of the field study allowing us to account for potential biases and confounds identified. The lab study procedure and questionnaires are based on a framework proposed, and validated, by Knijnenburg et al. \cite{Knijnenburg2012-hf} for evaluating recommender systems using a user-centric approach, an overview of the framework is presented in Figure~\ref{fig_framework}. Ethical approval was given for the lab study by the Swansea University Faculty of Science and Engineering Ethics Committee (SU-Ethics-Student-140323/6255).

\subsection{Procedure}
The study took place in the User Experience Lab and Legacy Lab in the Computational Foundry building at Swansea University. During the study, only the researcher and a single participant were in the room. Consent was given by participants prior to any study related questions.

\subsubsection{The Modified Genetic Car Designer}
In the lab study participants completed their task using a modified version of the genetic car designer. Modifications were made to account for the smaller sample size and to control the potential biases and confounds identified in the field study. The modifications are explained below.

\paragraph{Only two gallery views were available, one generated using MAP--Elites and the other containing random designs as a control.} Limiting the choice to two galleries reduces the study to an A/B test more suited to a smaller sample size.

\paragraph{The gallery views were labelled "Insights 1" and "Insights 2"} to eliminate any bias introduced by the name. Galleries were randomly labelled in a double-blind fashion, and which insight corresponded to which gallery for which participant was only revealed during data analysis. Although in the large scale study we found no gallery was more likely to be selected first than the others, the gallery participants interacted the most with was the Speed Insights. It might be the case that since the task was to travel as far as possible participants would be naturally drawn to the Speed Insights.

\paragraph{The Geometry Insights gallery was selected as the MAP--Elites generated view.} Since each view populates its list of selections using a different algorithm the number of unique designs each view presents to the user over a session is different. This different number of unique designs may introduce a confound when comparing the number of selections from each view. In the large scale study there was the smallest difference between the number of selections from the Geometry view and the control, therefore minimising the likelihood any differences we may observe are due to the confound identified.

\paragraph{The course, number of degrees of freedom, and random seed were fixed for all participants.} Fixing these variables ensured that the task difficulty was uniform across participants. Furthermore, fixing the random seed ensured the first set of designs was the same for all participants eliminating the confound between initial design quality and percentage improvement over the design session. 

\paragraph{Participants were given a fixed end point for the task, which was when the evolutionary algorithm has completed 40 generations.} In the large scale study we found that longer design sessions were more effective, in terms of percentage improvement over the session, and therefore time is a variable we need to control.

\subsubsection{Participants}
Participants were recruited from the student population of Swansea University under the following two criteria:
\begin{enumerate}
\item The participant had not used The Genetic Car Designer before (ensuring uniform situational characteristics (SC) among participants \cite{Knijnenburg2012-hf}), and
\item the participant had completed or is working towards an undergraduate degree in engineering and therefore familiar and practised in the concept of engineering design (to reduce variability of personal characteristics (PC)).
\end{enumerate}
Demographic information was collected to inform us of PC~\cite{Knijnenburg2012-hf} of participants. A total of 12 participants were recruited, 8 of whom self identified as male, 3 female and 1 non-binary. At the time of the study 8 of the participants were postgraduate students and 4 undergraduates. The mean age of participants was $24$ ranging between $20$ and $28$. Participants were remunerated for their time and given a £20 shopping voucher; the rate of remuneration was in line with the policies of our institute.

\paragraph{Pre-task Survey} \label{sec_pre_task}
In addition to the demographic questions, participants were asked a series of questions to gauge their attitudes towards technology, which further informs us of participants' personal characteristics. At this stage in the experiment the participants are unaware of the task and tool. Table~\ref{tab_pre_task_response} shows the participant responses to the pre-task survey. Generally the group of participants is pro-technology and confident using technology, with the main differences of opinions in the group related to trusting technology and preferring to do calculations by hand.
% \begin{table*}[bt!]
% \caption{Participant responses for the pre-task personal characteristic questions. In the table A, I and D refer to Agree, Indifferent and Disagree respectively. }\label{tab_pre_task_response}
% \begin{tabularx}{\linewidth}{X | l l l l l l l l l l l l}
% \toprule
%  & A & B & C & D & E & F & G & H & I & J & K & L \\
%  \midrule
% I prefer to do calculations by hand. & A & D & D & A & D & D & D & A & A & D & D & A\\
% I have no problem trusting my life to technology. & A & A & D & I & A & I & I & D & I & A & I & A\\
% I always double-check results from automated methods. & D & A & A & A & A & D & A & D & I & A & I & A\\
% I am confident using most technology. & D & A & A & A & I & A & I & A & A & A & A & D  \\
% The usefulness of technology is overrated. & I & D & D & D & D & D & D & D & D & D & D & I\\
% \bottomrule
% \end{tabularx}
% \end{table*}

\begin{table}
\caption{Participant responses for the pre-task personal characteristic statements.} \label{tab_pre_task_response}
\begin{tabularx}{\linewidth}{X | l l l l l l l l l l l l}
\toprule
Agree: \cmark \; Disagree: \xmark \; Indifferent: -- & A & B & C & D & E & F & G & H & I & J & K & L \\
 \midrule
I prefer to do calculations by hand. & \cmark & \xmark & \xmark & \cmark & \xmark & \xmark & \xmark & \cmark & \cmark & \xmark & \xmark & \cmark\\
I have no problem trusting my life to technology. & \cmark & \cmark & \xmark & -- & \cmark & -- & -- & \xmark & -- & \cmark & -- & \cmark\\
I always double-check results from automated methods. & \xmark & \cmark & \cmark & \cmark & \cmark & \xmark & \cmark & \xmark & -- & \cmark & -- & \cmark\\
I am confident using most technology. & \xmark & \cmark & \cmark & \cmark & -- & \cmark & -- & \cmark & \cmark & \cmark & \cmark & \xmark  \\
The usefulness of technology is overrated. & -- & \xmark & \xmark & \xmark & \xmark & \xmark & \xmark & \xmark & \xmark & \xmark & \xmark & --\\
\bottomrule
\end{tabularx}
\end{table}

\subsubsection{Design Task}
An explanation of the Genetic Car Designer and task was given to participants including:
\begin{itemize}
    \item The design goal;
    \item How to start the software;
    \item How to interact with each feature in the tool;
    \item The duration they have to complete the task; and
    \item What to do if they get stuck or have a question.
\end{itemize}
At no point during this explanation is the difference between the two insights views discussed or revealed. The insight views were described as lists of previous designs that had been simulated during the design session which could be interacted with in a similar way to the live view. After the explanation participants started the task, during the task participants were encouraged to use all the tools available to them. During the study the same analytical data collected in the large scale study was recorded, which enables us to evaluate objective aspects~\cite{Knijnenburg2012-hf} of the system.

\subsubsection{Post-task Survey}
Once the task was complete the participants were given two sets of questions. The first set is presented in Table~\ref{tab_AB_Statements}, these are designed to evaluate the subjective system aspects \cite{Knijnenburg2012-hf} by asking participants to compare Insights 1 to Insights 2.
\begin{table}
    \caption{Questions presented to participants to complete post task. Participants could answer Insights 1, Insights 2 or Neither. Questions are presented in the order they were given to participants.}\label{tab_AB_Statements}
    \begin{tabularx}{\linewidth}{l X}
    \toprule
    {Aspect} & {Statement}  \\
    \midrule    
    Subjective Accuracy & Which presented better solutions? \\
     & Which allowed you to select more optimal recommendations? \\
     & Which showed too many poor recommendations?  \\
    \midrule 
    Subjective Diversity & Which presented a variety of choices?  \\
     & Which showed a bigger difference in recommendations?  \\
     \midrule 
    Subjective Satisfaction & Which list was more valuable with respect to your time?  \\
     & Which had more satisfying recommendations?  \\
     & Which would you trust more to provide you with recommendations? \\
     & Which made finding a new solution easier?  \\
     \midrule
     Subjective Novelty & Which gave you more recommendations you would not expect? \\
     & Which gave you recommendations you would not have thought of yourself? \\
     & Which allowed you to explore new ideas better? \\
    \bottomrule
    \end{tabularx}
\end{table}
Finally a set of open-ended questions were given to participants to evaluate the experience and interaction aspects of the tool \cite{Knijnenburg2012-hf}, these questions are presented in Table~\ref{tab_questions}.
\begin{table}
    \caption{Open ended questions given to participants at the end of the lab study. Questions are presented in the order they were given to participants.}\label{tab_questions}
    \begin{tabularx}{\linewidth}{l X}
    \toprule
    {Aspect} & {Question}  \\
    \midrule    
    Interaction & How easy or difficult was the car designer to use? \\
    Experience & How easy or difficult was it to make a decision from the lists?  \\
    Experience & Was the decision process frustrating for insights 1 or 2?  \\ 
    Interaction \& Experience & How much effort did you need to invest while using insights 1 and 2?  \\
    Experience & Do you trust the results from the car designer?  \\     
    \bottomrule
    \end{tabularx}
\end{table}

\subsection{Results}
%Should there be a small paragraph here?
\subsubsection{Objective System Aspects}
\begin{figure}
\centering
\includegraphics[width=.9\textwidth]{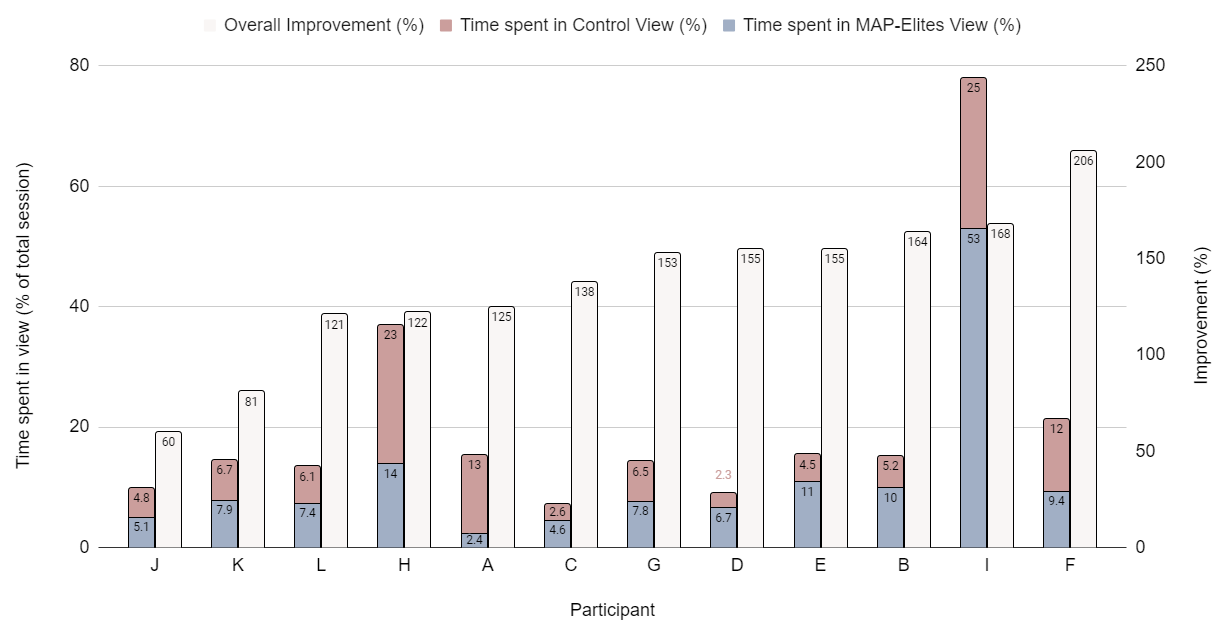}%
\caption{A stacked bar chart showing the time participants spend in the MAP--Elites (blue) and Control (red) galleries alongside the overall improvement (white). Participants are sorted by overall improvement.}
\label{fig_lab_study_views_time}
\end{figure}

\begin{figure}
\centering
\includegraphics[width=.9\textwidth]{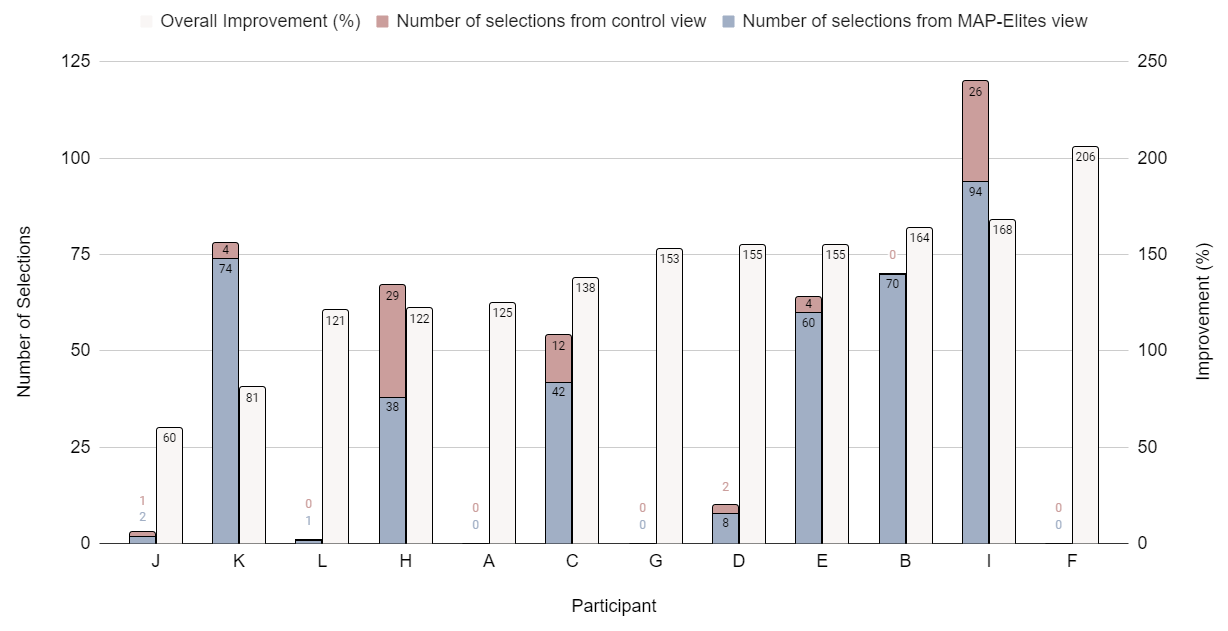}%
\caption{A stacked bar chart showing number of selections from the MAP--Elites gallery (blue), the Control gallery (red) alongside overall improvement (white). Participants are sorted by overall improvement.}
\label{fig_lab_study_views_selections}
\end{figure}

\paragraph{Most participants spent more time viewing and selecting designs from the MAP--Elites gallery, but there was no significant correlation between this level of cognitive and behavioural engagement with the galleries and design quality.} Figures~\ref{fig_lab_study_views_time} and~\ref{fig_lab_study_views_selections} respectively show the time each participant spent in the control and MAP--Elites gallery views and the number of selections they made from the two galleries. The overall performance in terms of improvement is also shown on these figures and participants are ranked by that performance. Most (75\%) participants engaged cognitively more with the MAP--Elites gallery compared to the Control gallery, and all participants who behaviourally engaged with at least one of the galleries selected more from the MAP--Elites view. Three participants (A, G and F) did not make any selections from either gallery and only interacted with the editor. No significant correlation between improvement and either time spent or selections from each gallery is observed.

\subsubsection{Subjective System Aspects}
\paragraph{Despite objective behavioural evidence suggesting that the MAP--Elites gallery was most useful, participants had overall mixed subjective opinions on the relative usefulness of the two galleries.} Table~\ref{tab_post_task_AB} shows the responses from participants for the post task comparison style questions. Using a Chi-Square Goodness of Fit Test for each question, with the null hypothesis that the frequency of responses is equal, we were unable to reject the null hypothesis at ($p<0.1$) for any question.  
\begin{table*}[bt!]
\caption{Participant responses for the post task comparison questions. In the table M-E, C and N refer to MAP--Elites, Control and Neither respectively. In the actual survey participants answered Insights 1 or Insights 2 these have been mapped to the correct view here.}\label{tab_post_task_AB}
\begin{tabularx}{\linewidth}{X | l l l l l l l l l l l l}
\toprule
 & A & B & C & D & E & F & G & H & I & J & K & L \\
 \midrule
Which presented better solutions? & N & M-E & C & N & C & M-E & M-E & C & M-E & M-E & C & M-E \\
Which allowed you to select more optimal recommendations? & C & M-E & C & C & C & C & M-E & C & M-E & M-E & M-E & M-E \\
Which showed too many poor recommendations? & M-E & C & N & M-E & M-E & C & C & M-E & C & C & N & C \\
\midrule
Which presented a variety of choices? & C & M-E & C & N & M-E & C & C & M-E & M-E & C & M-E & C \\
Which showed a bigger difference in recommendations? & N & M-E & C & M-E & M-E & C & C & M-E & C & C & C & M-E \\
\midrule
Which was more valuable with respect to your time? & N & M-E & C & C & C & C & M-E & N & M-E & M-E & N & M-E \\
Which had more satisfying recommendations? & C & M-E & C & C & C & N & M-E & M-E & M-E & M-E & N & M-E \\
Which would you trust more to provide you with recommendations? & C & M-E & C & C & C & N & M-E & C & M-E & M-E & N & M-E \\
Which made finding a new solution easier? & C & M-E & C & M-E & C & C & M-E & C & M-E & M-E & C & M-E \\
\midrule
Which gave you more recommendations you would not expect? & M-E & M-E & M-E & C & M-E & C & C & C & C & C & M-E & M-E \\
Which gave you recommendations you would not have thought of yourself? & C & M-E & N & M-E & M-E & C & C & N & M-E & M-E & C & N \\
Which allowed you to explore new ideas better? & C & M-E & N & C & M-E & C & M-E & C & M-E & M-E & M-E & M-E \\
\bottomrule
\end{tabularx}
\end{table*}
The responses show a mix of opinions regarding subjective performance of the two views and allow us to identify three groups of participants:
\begin{itemize}
    \item Participants who considered the MAP--Elites view to perform better overall
    \item Participants who thought the Control view performed better overall
    \item Participants who had mixed opinions on the subjective performance
\end{itemize}

%Currently separated
These feelings give insight into participants emotional engagement. Participants were placed in relevant groups if a majority of responses indicated a positive opinion of the relevant view, those participants who did not have a clear majority opinion were then placed in the mixed group. The groupings of participants are shown in Tables~\ref{tab_post_group_time} and~\ref{tab_post_group_selection} alongside their objective behaviour with respect to which views they engaged with the most cognitively and behaviourally. 
\begin{table}
\caption{Table showing the grouping of participants based on their opinions' of subjective performance of the views and their objective time behaviour. Participants who prefer to do calculations by hand are written in {\bf{bold}}.}\label{tab_post_group_time}
\begin{tabularx}{\textwidth}{X | X X}
\toprule
& Participants who spent more time in the MAP--Elites view & Participants who spent more time in the Control view\\
\midrule
Participants who considered the MAP--Elites view to perform better overall. & B, {\bf{I}}, J, {\bf{L}} & ---\\
Participants who thought the Control view performed better overall. &  C, {\bf{D}}  & {\bf{A}}, F, {\bf{H}}\\
Participants who had mixed opinions on the subjective performance. & E, G, K & ---\\
\bottomrule
\end{tabularx}
\end{table}
\begin{table}
\caption{Table showing the grouping of participants based on their opinions' of subjective performance of the views and their objective selection behaviour. Participants who prefer to do calculations by hand are written in {\bf{bold}}.}\label{tab_post_group_selection}
\begin{tabularx}{\linewidth}{X | X X}
\toprule
& Participants who selected more suggestions from the MAP--Elites view & Participants who did not select any suggestions from either view\\
\midrule
Participants who considered the MAP--Elites view to perform better overall. & B, {\bf{I}}, J, {\bf{L}} & ---\\
Participants who thought the Control view performed better overall. &  C, {\bf{D}}, {\bf{H}}  & {\bf{A}}, F \\
Participants who had mixed opinions on the subjective performance. & E, K & G\\
\bottomrule
\end{tabularx}
\end{table}

\paragraph{33\% of participants perceived MAP--Elites gallery to be more effective overall and this perception matched their behaviour in terms of time and number of selections.} Participants B and I gave short responses to the free text questions with Participant I simply describing the task as easy, and Participant B saying most of the designs performed well "without any input from me". Participant J was "not clear how [either insight view] were affecting the car design" and had to put effort (cognitive engagement) into "figuring out how the insights were helping". Participant L found the insights useful to compare their own designs to "better performing cars" noting that they found insights more useful towards the end of the task "once the edits I made were no longer yielding better results".

\paragraph{25\% of participants perceived the Control gallery to be more effective overall but this did not match their behaviour in terms of number of selections.} Participant C found that the "numerous designs" presented by the insight views made it easy to select designs with potential viability.  Participant D described their approach as "experimental" with their focus being "more on what was going on in the live screen than insights". Participant H found that the insights worked well "in tandem with one another" and picked poor performing designs to explore new ideas, they also noted that their own preconceptions made the task challenging to begin with. 

\paragraph{25\% of participants had mixed subjective opinions on the effectiveness of the two views.} Participant E only discussed how the fitness values displayed assisted decision making and did not refer to either set of insights. Participant K had a similar approach of using the fitness values as the primary support for decision making, adding that "it wasn't obvious what insights 1 and 2 represented". Participant G however was the only participant to directly compare the two insight views saying that "[Control] seemed more random and messy. [MAP--Elites] less choices but seemed better designs". Interestingly, participants E and K used selections from the MAP--Elites insights view more than most other participants despite their mixed views.

\paragraph{17\% of participants perceived the Control to be more effective overall, did not make any selections from either view, but did spend most time in the Control view.} Participant A focused on optimising a single design, finding that some of the designs in the insights were much better than the rest, which helped make decision making less frustrating. However, Participant A noted "there was limited information on what I was actually looking at [in the insights]". Participant F found that the insights "provided variations on the designs I was using to copy" and that both insights provided a series of "visually distinct" variations.

\subsubsection{Qualitative Analysis of Interaction and Experience}
We used Qualitative Content Analysis~\cite{Mayring2015-hm} to analyse participants' free-text responses, as the structured nature of our questions lent themselves to a deductive approach. SW and JV independently reviewed the data and proposed initial coding categories. After discussing and refining these, they agreed on a final codebook (Table~\ref{tab_codes}). Both then coded the data independently, resolving discrepancies through discussion until full consensus was reached, with the coded responses shown in Table~\ref{tab_code_results}.
\begin{table}
    \caption{Codebook used for analysing responses to the free text questions asked post-task.}\label{tab_codes}
    \begin{tabularx}{\linewidth}{>{\hsize=.05\hsize}X >{\hsize=.1\hsize}X >{\hsize=.15\hsize}X >{\hsize=.7\hsize}X}

    %|>{\hsize=.5\hsize}X|>{\hsize=.25\hsize}X|>{\hsize=.25\hsize}X|
    
    \toprule
    {ID} & {Category} & {Code} & {Definition}  \\
    \midrule    
    D1 & Difficulty & Easy& A participant found one or more aspects of the car designer easy to use.\\
D2 & Difficulty & Neutral difficulty & A participant found it was neither easy nor difficult to use one or more aspects of the car designer.\\
D3 & Difficulty  & Difficult & A participant found one or more aspects of the car designer difficult to use.                                        \\
D4 & Difficulty  & Easier with time & A participant found one or more aspects of the car designer easier to use as time went on.\\
\midrule
F1 & Frustration & Not frustrating & A participant did not find one or more aspects of the car designer frustrating. \\
F2 & Frustration & Neutral frustration & A participant was somewhat frustrated with one or more aspects of the car designer. \\
F3 & Frustration & Frustrating & A participant did find one or more aspects of the car designer frustrating. \\
F4 & Frustration & Equal & A participant found the same level of frustration with both insights. \\
\midrule
E1 & Effort & Minimal effort & A participant used little to no effort when using one or more aspects of the car designer. \\
E2 & Effort & Some effort & A participant used some effort when using one or more aspects of the car designer. \\
E3 & Effort & Lots of effort & A participant used lots of effort when using one or more aspects of the car designer. \\
\midrule
T1 & Trust & Trust & A participant trusted the results of the car designer. \\
T2 & Trust & Neutral trust & A participant trusted some of the results of the car designer, but had reservations. \\
T3 & Trust & Distrust & A participant did not trust the results of the car designer. \\
\midrule
S1 & Design & Seeing + Understanding & A participant expressed interest in finding the correlation between changing parameters and the car's performance. \\
S2 & Design & Preconceptions & A participant's preconceptions were mentioned as an influence on the design process. \\
S3 & Design & Score based & A participant used the scores provided to direct their design process. \\
S4 & Design & Vision based & A participant used visual information to direct their design process. \\
S5 & Design & Time & A participant mentioned time as a motivating factor in the design process. \\
\midrule
U1 & UI & Sliders & A participant mentioned aspects of interacting with the editor. \\
U2 & UI & Clutter & A participant mentioned separating out cars while the car designer was running. \\
\midrule
I1 & Insights & Insights unknown & A participant was confused about the role of insights in the design process or how to use it. \\
I2 & Insights & Insights known & A participant identified one or more characteristics of the insights and/or used the insights in the design process. \\
    \bottomrule
    \end{tabularx}
\end{table}
\begin{table}
\caption{Coded responses from each participant for the post task free response questions.}\label{tab_code_results}
\newcolumntype{s}{>{\hsize=0.05\hsize}X}
\begin{tabularx}{\linewidth}{>{\hsize=.3\hsize}X | s s s s s s s s s s s s}
\toprule
 & A & B & C & D & E & F & G & H & I & J & K & L \\
 \midrule
How easy or difficult was the car designer to use? & D1, D4& D4 & D1 & D1, U2 & D4, U1, S1 & D1, U1 & D1, U1 & D3, S2 & D4 & D1 & D1, U1, I1, U2 & D4, U2, U1 \\
\midrule
How easy or difficult was it to make a decision from Insights 1 and 2? & D3, S3 & D1, S4 & D1 & D3 & D2, S3, S2 & D1, I2 & D1, S3, I2 & S2, S3, S4 & D1 & I1 & S3 & D1, S3 \\
\midrule
Was the decision process frustrating for Insights 1 or 2? & F2, S3 & F1 & F1 & F1 & F1 & F1, I2 & F1, S3, S4 & F1, I2 & F1 & F1 & F4 & F2, S5 \\
\midrule
How much effort did you need to invest while using Insights 1 or 2? & E3, I1 & E1 & E1, I2 & S4 & E1, S3 & E1, S3 & I2, S5, U2 & E1, S3 & E1 & E2, I1 & E1, S3 & E2 \\
\midrule
Do you trust the results from the car designer? & T3, S1 & T1 & T1, S4 & T1 & T1 & T1, S3 & T1, S3, S4 & T1 & T1 & T2 & T1, S2 & T2, S1 \\
\bottomrule
\end{tabularx}
\end{table}

%Do this based on categories in the code table
\emph{Most (58\%) participants found the genetic car designer easy to use and some (42\%) found it got easier to use the longer they used it.} Only Participant H described the designer difficult to use, saying that ``finding an optimal design was harder than expected''. Six participants related ease of use to the user interface (UI). For example, Participant G ``Struggled to figure out which numbers corresponded to which part geometry at the start'' and Participant K said ``There were a lot of cars so it wasn't clear where the car you had edited was''.

\emph{The majority (75\%) of participants reported that the decision process was not made frustrating by the insights.} The main frustration point, which was raised by Participant L, was that the gallery ``list changed so regularly''. Similarly, Participant A found the system ``A little difficult as insight recommendations kept changing''.

\emph{Half the participants described the effort needed to use the insights to be minimal}, many of those (4 out of 6) relating this lack of effort to simply selecting the design with the best score. For example, Participant H states ``Not much effort was used, in the process of evaluating using the insights. Simply looking for the highest values.''. Only Participant A described the amount of effort as high saying ``More effort than what I expected to need. There was limited information on what I was actually looking at.''. Some participants gave some detail on their understanding of what the insights were. For example, Participant F stated that the insights helped ``as they provided variations on the designs I was using to copy the success of'' and Participant J was unsure ``how the insights 1 or 2 were affecting the car design''.

\emph{Most (75\%) of the participants reported that they trusted the results from the car designer.} Reasons for trusting the designer largely were related to the performance of designs. For example, Participant C said ``I do, there were a lot of viable designs generated that with a little tweaking performed very well.'' and ``The final designs (about 3 different configurations) were consistently the highest scoring, and looked the most sensible''. Conversely Participant A was the only participant to state that they did not trust the designer because ``although some changes made sense, I couldn't easily rely on a certain configuration or change I made''.

\emph{Most (58\%) participants reported focusing on the numerical value in their design approach, whereas two participants reported focusing on the visual aspects of design.} For example, Participant E found ``The scoring system helped make the decisions much easier.'', conversely Participant H stated ``I tended to use my instinct for which design aligned best with my idea for what it should look like.''.

\section{Discussion}
%The Discussion section should be used to emphasise the new and important aspects of the study, placing the results in context with published literature and the implications of the findings.

\subsection{RQ1: Do galleries of examples influence user engagement with the design process?}
A significant, and striking, result from the field study was that participants who viewed at least one gallery were more cognitively engaged in the design task compared to those who did not. The average session length for participants who viewed at least one gallery was more than twice that of participants who did not. This finding adds significant strength to existing claims~\cite{Walton2021-vi, Mo2024-ku}, based on much smaller studies than ours, that algorithm-generated galleries of examples increase engagement in design tasks. Our conclusion is built on the assumption that the length of a session (in our field study) is a measure of cognitive engagement, we believe this is a fair assumption, but recognise that rigorous small-scale investigations into engagement are required.

Participants who interacted with at least one gallery in the field study, by selecting designs to offer as suggestions to the evolutionary algorithm, spent the longest on the task. This indicates a strong relationship between behavioural and cognitive engagement which is challenging to untangle due to limitations of our approach. In future work we recommend taking care in designing a methodology to isolate these two aspects of engagement by, for example, using methods such as eye tracking as a proxy for cognitive engagement.

Throughout the responses from participants in the lab study there is evidence of emotional engagement, in terms of positive and negative feelings, with the galleries and task. Using these responses to assess participant's emotional engagement with specific galleries we found that their opinions and feelings did not always match their behaviour in terms of actions. Only 33\% of participants in the lab study expressed positive emotions towards the gallery they engaged the most with behaviourally and cognitively. This suggests that behavioural and cognitive engagement are not predicted by positive emotional engagement, and vice versa, which is important to consider when designing studies to evaluate engagement in design tasks.

\subsection{RQ2: Do galleries of examples have an influence on the quality of designs produced?}
%This question warrants more investigation
Our results show that increased cognitive engagement in the design task resulting from galleries of examples leads to longer design sessions and, hence, better quality designs. The results from our field study showed that sessions where participants selected designs from the galleries to be entered into the breeding and elite pools resulted in the highest improvement of design quality. In other words, higher levels of behavioural engagement resulted in better quality designs.  Furthermore, although this difference in correlation was not statistically significant, we observed that the negative correlation between initial design quality and overall improvement was lowest in sessions where participants interacted with at least one gallery. These results could be entirely algorithmic in nature since we would expect that providing the optimiser better designs to generate the next generation could lead to improved performance. What is interesting, and initially indicates an effect beyond the algorithm, is that the act of viewing a gallery resulted in increased performance. However, our results also show that the amount of time spent in a session was positively correlated to the quality of designs produced, and simultaneously that participants who viewed galleries spent longer on the task.
%However, our results also show that the longer a session the better the quality of designs produced, and simultaneously that participants who viewed galleries spent longer on the task. 

In the lab study, which had a fixed session length, we found no correlation between design quality and amount of behavioural engagement with the galleries. It follows that the positive correlations observed between amount of engagement with galleries and quality of design observed in the field study could be due to the increased cognitive engagement in the task, leading to longer design sessions, which gives the algorithm and human more time to improve designs. In our lab study, participants did report that the recommendations were improving the quality of their designs, evidence of emotional engagement. When asked to pick the best insight view based on the quality/accuracy of recommendations only 3 of the 36 responses were "Neither". Further to this, many participants detailed that the insight views were helping them in the design process in free text responses.

\subsection{RQ3: Do galleries of examples generated using MAP--Elites have a different utility compared to a random set of suggestions?}
Our results show that there is indeed a difference in utility when comparing MAP--Elites driven galleries to the random control and that neither is necessarily more valuable than the other to the design process. The results from our field study show that participants selected significantly more designs from the MAP--Elites driven galleries for inclusion in the elite and breeding pools of the evolutionary algorithm compared with the random control. This finding holds despite participants not spending significantly more time viewing MAP--Elites driven galleries. This finding carries over to our lab study where we found that all participants selected more designs from the MAP--Elites driven gallery compared with the control. Many participants from the lab study noted that the MAP--Elites gallery provided better quality designs than the control. When asked their subjective opinions on the relative usefulness of the MAP--Elites and random control galleries in the design process, neither gallery was highlighted more than the other by participants. Interestingly, 25\% of participants reported that the random control was the most useful gallery despite making more selections from the MAP--Elites gallery, indicating that number of selections from a gallery (i.e. behavioural engagement) is not a good measure for utility. The range of perceptions observed during the lab study is potentially due to the varying approaches detailed by the participants. Some participants explained how they took an experimental or exploratory approach to the task, not simply focusing on the best performing designs but also gaining understanding by identifying poor performing designs. Several participants, who perceived the MAP--Elites view as more effective, expressed a desire to understand the effect each degree of freedom had on performance, indicating that they were building a mental model of the system (cognitive engagement). Participants who, conversely, perceived the control insight as being effective tended to report a more exploratory approach compared to those who perceived the MAP--Elites insights as more effective. This suggests that the structured diversity provided by the MAP--Elites galleries supports a more systematic exploration of the design space than a non-intelligent system. Interestingly, the participants who focussed solely on the fitness scores of each design to make decisions reported making more use of the MAP--Elites view, suggesting that this gallery facilitated a more satisfying user experience. The only difference between the MAP--Elites gallery and the random control is the diversity of suggestions, based on our results we posit that diversity of suggestions is a crucial aspect perhaps enhancing the user's perception of informativeness or reducing decision fatigue through visual diversity. In conclusion, MAP--Elites should not be considered as a back-end tool for design optimisation but instead core part of the user experience in computational design.

\subsection{Limitations of our Approach}
A key aim with our approach was to attract a large pool of participants and therefore increase the statistical power of our findings. The cost of designing a study which is approachable by a wide audience is that our findings may not generalise to engineers in a professional setting. To mitigate this limitation we recruited student engineers, with design training, for our small scale lab study. However, clearly it would be useful to carry out a focused study with design engineers within a professional setting. It is our hope that the findings and recommendations we present here can be used to design an effective study involving engineering professionals. 

A number of participants in the lab study discussed issues with the user experience (UX) of the design tool which may have effected our findings. For example, a number of participants expressed that the insight views were difficult to use because they updated too frequently. This highlights the importance of UX design which is often overlooked when designing research software. In future studies care should be taken to ensure the UX/UI does not get in the way of testing the core algorithms which are part of the research questions.

One limitation of our approach is that it is challenging to separate the effects of the control gallery and MAP--Elites gallery views. Many participants in our lab study expressed that they used multiple galleries in tandem with one another. This raises questions of whether or not the effectiveness of the different algorithms used to generate galleries would change if they were considered in isolation. A study is needed where participants only view galleries generated by a single algorithm to answer this question of dependence.

\subsection{Implications and Future Work}
The results of our studies provide strong and compelling evidence that galleries of example designs enhance both user engagement and the quality of designs produced in human--AI collaborative environments. Our findings align and add strength to previous research highlighting the benefits of mixed-initiative tools in fostering creativity and improving design outcomes~\cite{Lee2010-wa, Ngoon2018-iu, Swearngin2020-xt}. In this subsection we detail key implications and directions for future work.

\subsubsection{Opportunities with Citizen Science}
In half the sessions recorded in our field study participants were completely passive and simply observed the algorithm or left it running. This provided a useful baseline for algorithm performance, but it is worthwhile considering what caused participants to be passive. After completing data collection we discovered that a small number of older browsers initially hid the view controls from view, which may explain some of the passive participants. We found that passive participants were more likely to select more complex problems with much higher numbers of degrees of freedom, this could be an indication that they were playing with the tool. It could highlight the growing interest in the general public towards algorithms which exhibit creativity~\cite{Ploin2022-ml} and may represent an opportunity to conduct more research using citizen science.

\subsubsection{Human--AI collaborative environments may not save time}
When answering RQ2 we found that galleries of examples led to increased engagement and longer design sessions, which yielded better design outcomes. There is an open question of what causes this effect, is it that giving the optimisation algorithm longer to solve the problem leads to better design quality, or that giving the human designer longer does the same, or a combination of both? We found a similar result in a previous study~\cite{Walton2021-vi} when comparing the use of an AI-based tool to a non-AI based to design five levels for a video game. To our surprise the participants who used the AI-based tool took significantly longer to complete the task, when digging into the qualitative data we found this was because their emotional engagement in the task was higher. Gallery based human--AI collaborative tools should not be seen as tools to save time and money~\cite{Melotti2019-rt, Ruela2018-ah, Baldwin2017-me} but instead tools which lead to better outcomes by enhancing human creativity.

\subsubsection{Evaluating gallery based Human--AI collaborative environments}
A key finding from both studies is that simply viewing suggestions in a gallery has a positive influence on the design process, it is therefore important that time viewing galleries is a key metric used when evaluating algorithms. In contrast, current assessment methodologies for mixed-initiative systems are typically based on measuring how often users select and edit suggestions from the algorithm~\cite{Yannakakis2014-rt, Walton2021-vi}, and as such can only evaluate behavioural engagement. Although responses from participants in the lab study confirm that viewing recommendations did have an effect on the design process, and supports our assumption that view time is a proxy for cognitive engagement, a limitation of our approach is a lack of eye tracking or other physical engagement measures. A future study including some form of eye tracking, to determine which specific design suggestions participants are looking at as they make decisions, would allow a more rigorous comparison between different methods for generating galleries. Furthermore, it may be fruitful to explore more modes of physical engagement measures, such as using electroencephalography devices~\cite{Yan2017-au}.

\subsubsection{Rethinking the evaluation of generative systems}
Whist our study does not directly consider generative AI, our findings have implications for the design and evaluation of generative systems. Notably, the observation we present in the introduction, that HCI and videos game research have historically approached human--AI collaboration from opposite directions---one emphasising supporting the design process, the other emphasising emergence and automation---becomes increasingly relevant as generative AI systems become more prolific in both domains. Future work may explore how structured diversity, afforded by techniques such as MAP--Elites, might complement or contrast the open-ended emergent nature of generative models. Convergence of these various areas invites consideration of new evaluation frameworks which account not only for task outcome quality, but also the engagement, agency, creativity and trust of the human user. For example, in video games research, significant effort has been placed into measuring and defining the experience of immersion~\cite{Jennett2008-pv} which is historically linked to engagement and may be relevant for human--AI system evaluation.

\subsubsection{Adaptive mixed-initiative algorithms}
In the lab study we found that depending on participants' approach to design they perceived the relative value of the control and MAP--Elites gallery differently. This suggests there may be some fruitful avenues in tailoring algorithms which generate suggestions based on the varying approaches of designers. As reported by many of our participants, their approaches changed throughout the design process so perhaps the algorithms which are used to generate suggestions also need to change as the design process continues. There has already been some work~\cite{Alvarez2020-xv} on techniques to learn preferences of designers during the design process which could be valuable here, allowing the AI to adapt as time progresses.  

\subsubsection{Encouraging engagement through trust}
The positive correlations between user engagement with the galleries and design quality underscores the need to build transparent and intuitive interfaces that encourages the human designer to engage with the AI. A number of participants in our lab study stated that they were trying, and in some cases failing, to understand what the AI was doing and what the gallery views meant. Arguments have been made~\cite{chanel2020mixed} that humans and AI should be seen as teammates for effective mixed-initiative systems, and for an AI to truly replicate a human collaborator it must be able to justify its decisions to build trust \cite{zhu2018explainable}. As we have previously found \cite{Vincalek2021-lz, Vincalek2021-lz2} the only way to increase the uptake of new methods within engineering is to make them trustworthy. Interestingly, some participants in our lab study linked trust in the algorithm to how much the designs it suggested matched their preconceptions, only trusting the algorithm when things "made sense".  Therefore, an important future avenue for work is to investigate how preconceptions affect the design process and how we can design tools to support designers to move beyond their preconceptions without breaking their trust.

%\section{Conclusion}

\section*{Acknowledgement}
This work was supported by EPSRC under Grant Engineering and Physical Sciences Research Council EPSRC:EP/S021892/1.

%%
%% The next two lines define the bibliography style to be used, and
%% the bibliography file.
\bibliographystyle{ACM-Reference-Format}
\bibliography{sample-base}

%%% -*-BibTeX-*-
%%% Do NOT edit. File created by BibTeX with style
%%% ACM-Reference-Format-Journals [18-Jan-2012].

\begin{thebibliography}{43}

%%% ====================================================================
%%% NOTE TO THE USER: you can override these defaults by providing
%%% customized versions of any of these macros before the \bibliography
%%% command.  Each of them MUST provide its own final punctuation,
%%% except for \shownote{}, \showDOI{}, and \showURL{}.  The latter two
%%% do not use final punctuation, in order to avoid confusing it with
%%% the Web address.
%%%
%%% To suppress output of a particular field, define its macro to expand
%%% to an empty string, or better, \unskip, like this:
%%%
%%% \newcommand{\showDOI}[1]{\unskip}   % LaTeX syntax
%%%
%%% \def \showDOI #1{\unskip}           % plain TeX syntax
%%%
%%% ====================================================================

\ifx \showCODEN    \undefined \def \showCODEN     #1{\unskip}     \fi
\ifx \showDOI      \undefined \def \showDOI       #1{#1}\fi
\ifx \showISBNx    \undefined \def \showISBNx     #1{\unskip}     \fi
\ifx \showISBNxiii \undefined \def \showISBNxiii  #1{\unskip}     \fi
\ifx \showISSN     \undefined \def \showISSN      #1{\unskip}     \fi
\ifx \showLCCN     \undefined \def \showLCCN      #1{\unskip}     \fi
\ifx \shownote     \undefined \def \shownote      #1{#1}          \fi
\ifx \showarticletitle \undefined \def \showarticletitle #1{#1}   \fi
\ifx \showURL      \undefined \def \showURL       {\relax}        \fi
% The following commands are used for tagged output and should be
% invisible to TeX
\providecommand\bibfield[2]{#2}
\providecommand\bibinfo[2]{#2}
\providecommand\natexlab[1]{#1}
\providecommand\showeprint[2][]{arXiv:#2}

\bibitem[Alvarez et~al\mbox{.}(2018)]%
        {Alvarez2018-am}
\bibfield{author}{\bibinfo{person}{Alberto Alvarez}, \bibinfo{person}{Steve Dahlskog}, \bibinfo{person}{Jose Font}, \bibinfo{person}{Johan Holmberg}, \bibinfo{person}{Chelsi Nolasco}, {and} \bibinfo{person}{Axel {\"O}sterman}.} \bibinfo{year}{2018}\natexlab{}.
\newblock \showarticletitle{Fostering creativity in the mixed-initiative evolutionary dungeon designer}. In \bibinfo{booktitle}{\emph{Proceedings of the 13th International Conference on the Foundations of Digital Games}} (Malm{\"o}, Sweden) \emph{(\bibinfo{series}{FDG '18}, \bibinfo{number}{Article 50})}. \bibinfo{publisher}{Association for Computing Machinery}, \bibinfo{address}{New York, NY, USA}, \bibinfo{pages}{1--8}.
\newblock


\bibitem[Alvarez et~al\mbox{.}(2019)]%
        {Alvarez2019-oc}
\bibfield{author}{\bibinfo{person}{A Alvarez}, \bibinfo{person}{S Dahlskog}, \bibinfo{person}{J Font}, {and} \bibinfo{person}{J Togelius}.} \bibinfo{year}{2019}\natexlab{}.
\newblock \showarticletitle{Empowering Quality Diversity in Dungeon Design with Interactive Constrained {MAP-Elites}}. In \bibinfo{booktitle}{\emph{2019 {IEEE} Conference on Games ({CoG})}}. \bibinfo{publisher}{ieeexplore.ieee.org}, \bibinfo{pages}{1--8}.
\newblock


\bibitem[Alvarez and Font(2020)]%
        {Alvarez2020-xv}
\bibfield{author}{\bibinfo{person}{Alberto Alvarez} {and} \bibinfo{person}{Jose Font}.} \bibinfo{year}{2020}\natexlab{}.
\newblock \showarticletitle{Learning the designer's preferences to drive evolution}. In \bibinfo{booktitle}{\emph{International Conference on the Applications of Evolutionary Computation (Part of {EvoStar})}}. \bibinfo{publisher}{Springer}, \bibinfo{pages}{431--445}.
\newblock


\bibitem[Alvarez et~al\mbox{.}(2021)]%
        {Alvarez2021-wr}
\bibfield{author}{\bibinfo{person}{Alberto Alvarez}, \bibinfo{person}{Jose Font}, \bibinfo{person}{Steve Dahlskog}, {and} \bibinfo{person}{Julian Togelius}.} \bibinfo{year}{2021}\natexlab{}.
\newblock \showarticletitle{Assessing the Effects of Interacting with {MAP-Elites}}.
\newblock \bibinfo{journal}{\emph{AIIDE}} \bibinfo{volume}{17}, \bibinfo{number}{1} (\bibinfo{date}{Oct.} \bibinfo{year}{2021}), \bibinfo{pages}{124--131}.
\newblock


\bibitem[Baldwin et~al\mbox{.}(2017)]%
        {Baldwin2017-me}
\bibfield{author}{\bibinfo{person}{A Baldwin}, \bibinfo{person}{S Dahlskog}, \bibinfo{person}{J~M Font}, {and} \bibinfo{person}{J Holmberg}.} \bibinfo{year}{2017}\natexlab{}.
\newblock \showarticletitle{Mixed-initiative procedural generation of dungeons using game design patterns}. In \bibinfo{booktitle}{\emph{2017 {IEEE} Conference on Computational Intelligence and Games ({CIG})}}. \bibinfo{pages}{25--32}.
\newblock


\bibitem[Chan et~al\mbox{.}(2022)]%
        {Chan2022-gh}
\bibfield{author}{\bibinfo{person}{Liwei Chan}, \bibinfo{person}{Yi-Chi Liao}, \bibinfo{person}{George~B Mo}, \bibinfo{person}{John~J Dudley}, \bibinfo{person}{Chun-Lien Cheng}, \bibinfo{person}{Per~Ola Kristensson}, {and} \bibinfo{person}{Antti Oulasvirta}.} \bibinfo{year}{2022}\natexlab{}.
\newblock \showarticletitle{Investigating positive and negative qualities of human-in-the-loop optimization for designing interaction techniques}. In \bibinfo{booktitle}{\emph{CHI Conference on Human Factors in Computing Systems}}. \bibinfo{publisher}{ACM}, \bibinfo{address}{New York, NY, USA}.
\newblock


\bibitem[Chanel et~al\mbox{.}(2020)]%
        {chanel2020mixed}
\bibfield{author}{\bibinfo{person}{Caroline~PC Chanel}, \bibinfo{person}{Rapha{\"e}lle~N Roy}, \bibinfo{person}{Nicolas Drougard}, {and} \bibinfo{person}{Fr{\'e}d{\'e}ric Dehais}.} \bibinfo{year}{2020}\natexlab{}.
\newblock \showarticletitle{Mixed-Initiative Human-Automated Agents Teaming: Towards a Flexible Cooperation Framework}. In \bibinfo{booktitle}{\emph{International Conference on Human-Computer Interaction}}. Springer, \bibinfo{pages}{117--133}.
\newblock


\bibitem[Charity et~al\mbox{.}(2022)]%
        {Charity2022-bl}
\bibfield{author}{\bibinfo{person}{M Charity}, \bibinfo{person}{Isha Dave}, \bibinfo{person}{Ahmed Khalifa}, {and} \bibinfo{person}{Julian Togelius}.} \bibinfo{year}{2022}\natexlab{}.
\newblock \showarticletitle{Baba is Y'all 2.0: Design and Investigation of a Collaborative {Mixed-Initiative} System}.
\newblock \bibinfo{journal}{\emph{IEEE Trans. Comput. Intell. AI Games}} (\bibinfo{year}{2022}), \bibinfo{pages}{1--15}.
\newblock


\bibitem[Cook(2017)]%
        {Cook2017-gh}
\bibfield{author}{\bibinfo{person}{Michael Cook}.} \bibinfo{year}{2017}\natexlab{}.
\newblock \showarticletitle{A Vision For Continuous Automated Game Design}. In \bibinfo{booktitle}{\emph{Proceedings of the 13th Experimental AI and Games Workshop}}. \bibinfo{publisher}{AIIDE}.
\newblock


\bibitem[Craveirinha and Roque(2015)]%
        {Craveirinha2015-hm}
\bibfield{author}{\bibinfo{person}{Rui Craveirinha} {and} \bibinfo{person}{Licinio Roque}.} \bibinfo{year}{2015}\natexlab{}.
\newblock \showarticletitle{Studying an {Author-Oriented} Approach to Procedural Content Generation through Participatory Design}. In \bibinfo{booktitle}{\emph{Entertainment Computing - {ICEC} 2015}}. \bibinfo{publisher}{Springer International Publishing}, \bibinfo{pages}{383--390}.
\newblock


\bibitem[Doherty and Doherty(2019)]%
        {Doherty2019-ee}
\bibfield{author}{\bibinfo{person}{Kevin Doherty} {and} \bibinfo{person}{Gavin Doherty}.} \bibinfo{year}{2019}\natexlab{}.
\newblock \showarticletitle{Engagement in {HCI}: Conception, theory and measurement}.
\newblock \bibinfo{journal}{\emph{ACM Comput. Surv.}} \bibinfo{volume}{51}, \bibinfo{number}{5} (\bibinfo{date}{Sept.} \bibinfo{year}{2019}), \bibinfo{pages}{1--39}.
\newblock


\bibitem[Duan et~al\mbox{.}(2024)]%
        {Duan2024-bi}
\bibfield{author}{\bibinfo{person}{Peitong Duan}, \bibinfo{person}{Jeremy Warner}, \bibinfo{person}{Yang Li}, {and} \bibinfo{person}{Bjoern Hartmann}.} \bibinfo{year}{2024}\natexlab{}.
\newblock \showarticletitle{Generating automatic feedback on {UI} mockups with large language models}. In \bibinfo{booktitle}{\emph{Proceedings of the CHI Conference on Human Factors in Computing Systems}}, Vol.~\bibinfo{volume}{4}. \bibinfo{publisher}{ACM}, \bibinfo{address}{New York, NY, USA}, \bibinfo{pages}{1--20}.
\newblock


\bibitem[Earle et~al\mbox{.}(2022)]%
        {Earle2022-pk}
\bibfield{author}{\bibinfo{person}{Sam Earle}, \bibinfo{person}{Justin Snider}, \bibinfo{person}{Matthew~C Fontaine}, \bibinfo{person}{Stefanos Nikolaidis}, {and} \bibinfo{person}{Julian Togelius}.} \bibinfo{year}{2022}\natexlab{}.
\newblock \showarticletitle{Illuminating diverse neural cellular automata for level generation}. In \bibinfo{booktitle}{\emph{Proceedings of the Genetic and Evolutionary Computation Conference}} (Boston, Massachusetts) \emph{(\bibinfo{series}{GECCO '22})}. \bibinfo{publisher}{Association for Computing Machinery}, \bibinfo{address}{New York, NY, USA}, \bibinfo{pages}{68--76}.
\newblock


\bibitem[Fontaine et~al\mbox{.}(2019)]%
        {Fontaine2019-fy}
\bibfield{author}{\bibinfo{person}{Matthew~C Fontaine}, \bibinfo{person}{Scott Lee}, \bibinfo{person}{L~B Soros}, \bibinfo{person}{Fernando De~Mesentier~Silva}, \bibinfo{person}{Julian Togelius}, {and} \bibinfo{person}{Amy~K Hoover}.} \bibinfo{year}{2019}\natexlab{}.
\newblock \showarticletitle{Mapping hearthstone deck spaces through {MAP-elites} with sliding boundaries}. In \bibinfo{booktitle}{\emph{Proceedings of the Genetic and Evolutionary Computation Conference}} (Prague, Czech Republic) \emph{(\bibinfo{series}{GECCO '19})}. \bibinfo{publisher}{Association for Computing Machinery}, \bibinfo{address}{New York, NY, USA}, \bibinfo{pages}{161--169}.
\newblock


\bibitem[Jennett et~al\mbox{.}(2008)]%
        {Jennett2008-pv}
\bibfield{author}{\bibinfo{person}{Charlene Jennett}, \bibinfo{person}{Anna~L Cox}, \bibinfo{person}{Paul Cairns}, \bibinfo{person}{Samira Dhoparee}, \bibinfo{person}{Andrew Epps}, \bibinfo{person}{Tim Tijs}, {and} \bibinfo{person}{Alison Walton}.} \bibinfo{year}{2008}\natexlab{}.
\newblock \showarticletitle{Measuring and defining the experience of immersion in games}.
\newblock \bibinfo{journal}{\emph{Int. J. Hum. Comput. Stud.}} \bibinfo{volume}{66}, \bibinfo{number}{9} (\bibinfo{date}{Sept.} \bibinfo{year}{2008}), \bibinfo{pages}{641--661}.
\newblock


\bibitem[Knijnenburg et~al\mbox{.}(2012)]%
        {Knijnenburg2012-hf}
\bibfield{author}{\bibinfo{person}{Bart~P Knijnenburg}, \bibinfo{person}{Martijn~C Willemsen}, \bibinfo{person}{Zeno Gantner}, \bibinfo{person}{Hakan Soncu}, {and} \bibinfo{person}{Chris Newell}.} \bibinfo{year}{2012}\natexlab{}.
\newblock \showarticletitle{Explaining the user experience of recommender systems}.
\newblock \bibinfo{journal}{\emph{User Model. User-adapt Interact.}} \bibinfo{volume}{22}, \bibinfo{number}{4} (\bibinfo{date}{Oct.} \bibinfo{year}{2012}), \bibinfo{pages}{441--504}.
\newblock


\bibitem[Lai et~al\mbox{.}(2022)]%
        {Lai2022-qv}
\bibfield{author}{\bibinfo{person}{Gorm Lai}, \bibinfo{person}{Frederic~Fol Leymarie}, {and} \bibinfo{person}{William Latham}.} \bibinfo{year}{2022}\natexlab{}.
\newblock \showarticletitle{On {Mixed-Initiative} Content Creation for Video Games}.
\newblock \bibinfo{journal}{\emph{IEEE Trans. Comput. Intell. AI Games}} \bibinfo{volume}{14}, \bibinfo{number}{4} (\bibinfo{date}{Dec.} \bibinfo{year}{2022}), \bibinfo{pages}{543--557}.
\newblock


\bibitem[Lee et~al\mbox{.}(2010)]%
        {Lee2010-wa}
\bibfield{author}{\bibinfo{person}{Brian Lee}, \bibinfo{person}{Savil Srivastava}, \bibinfo{person}{Ranjitha Kumar}, \bibinfo{person}{Ronen Brafman}, {and} \bibinfo{person}{Scott~R Klemmer}.} \bibinfo{year}{2010}\natexlab{}.
\newblock \showarticletitle{Designing with interactive example galleries}. In \bibinfo{booktitle}{\emph{Proceedings of the SIGCHI Conference on Human Factors in Computing Systems}}. \bibinfo{publisher}{ACM}, \bibinfo{address}{New York, NY, USA}.
\newblock


\bibitem[Lee et~al\mbox{.}(2020)]%
        {Lee2020-db}
\bibfield{author}{\bibinfo{person}{Chunggi Lee}, \bibinfo{person}{Sanghoon Kim}, \bibinfo{person}{Dongyun Han}, \bibinfo{person}{Hongjun Yang}, \bibinfo{person}{Young-Woo Park}, \bibinfo{person}{Bum~Chul Kwon}, {and} \bibinfo{person}{Sungahn Ko}.} \bibinfo{year}{2020}\natexlab{}.
\newblock \showarticletitle{{GUIComp}: A {GUI} design assistant with real-time, multi-faceted feedback}. In \bibinfo{booktitle}{\emph{Proceedings of the 2020 CHI Conference on Human Factors in Computing Systems}}. \bibinfo{publisher}{ACM}, \bibinfo{address}{New York, NY, USA}.
\newblock


\bibitem[Liapis et~al\mbox{.}(2013)]%
        {Liapis2013-at}
\bibfield{author}{\bibinfo{person}{Antonios Liapis}, \bibinfo{person}{Georgios~N Yannakakis}, {and} \bibinfo{person}{Julian Togelius}.} \bibinfo{year}{2013}\natexlab{}.
\newblock \showarticletitle{Sentient Sketchbook: Computer-aided game level authoring}. In \bibinfo{booktitle}{\emph{{FDG}}}. \bibinfo{pages}{213--220}.
\newblock


\bibitem[Louie et~al\mbox{.}(2020)]%
        {Louie2020-gl}
\bibfield{author}{\bibinfo{person}{Ryan Louie}, \bibinfo{person}{Andy Coenen}, \bibinfo{person}{Cheng~Zhi Huang}, \bibinfo{person}{Michael Terry}, {and} \bibinfo{person}{Carrie~J Cai}.} \bibinfo{year}{2020}\natexlab{}.
\newblock \showarticletitle{Novice-{AI} music co-creation via {AI}-steering tools for deep generative models}. In \bibinfo{booktitle}{\emph{Proceedings of the 2020 CHI Conference on Human Factors in Computing Systems}}. \bibinfo{publisher}{ACM}, \bibinfo{address}{New York, NY, USA}.
\newblock


\bibitem[Lu et~al\mbox{.}(2017)]%
        {Lu2017-ll}
\bibfield{author}{\bibinfo{person}{Jean Hsiang-Chun Lu}, \bibinfo{person}{William~Roshan Quadros}, {and} \bibinfo{person}{Kenji Shimada}.} \bibinfo{year}{2017}\natexlab{}.
\newblock \showarticletitle{Evaluation of user-guided semi-automatic decomposition tool for hexahedral mesh generation}.
\newblock \bibinfo{journal}{\emph{Finite Elem. Anal. Des.}} \bibinfo{volume}{4}, \bibinfo{number}{4} (\bibinfo{date}{Oct.} \bibinfo{year}{2017}), \bibinfo{pages}{330--338}.
\newblock


\bibitem[Mahfoud(1996)]%
        {Mahfoud1996-qn}
\bibfield{author}{\bibinfo{person}{Samir~W Mahfoud}.} \bibinfo{year}{1996}\natexlab{}.
\newblock \emph{\bibinfo{title}{Niching methods for genetic algorithms}}.
\newblock \bibinfo{thesistype}{Ph.\,D. Dissertation}. \bibinfo{address}{USA}.
\newblock


\bibitem[Matsunaga(2020)]%
        {Matsunaga_undated-su}
\bibfield{author}{\bibinfo{person}{Rafael Matsunaga}.} \bibinfo{year}{2020}\natexlab{}.
\newblock \showarticletitle{HTML5 Genetic Cars: A genetic algorithm car evolver in {HTML5} canvas}.
\newblock \bibinfo{journal}{\emph{GitHub repository}} (\bibinfo{year}{2020}).
\newblock
\urldef\tempurl%
\url{https://github.com/red42/HTML5\_Genetic\_Cars}
\showURL{%
\tempurl}


\bibitem[Mayring(2015)]%
        {Mayring2015-hm}
\bibfield{author}{\bibinfo{person}{Philipp Mayring}.} \bibinfo{year}{2015}\natexlab{}.
\newblock \showarticletitle{Qualitative Content Analysis: Theoretical Background and Procedures}.
\newblock In \bibinfo{booktitle}{\emph{Approaches to Qualitative Research in Mathematics Education: Examples of Methodology and Methods}}, \bibfield{editor}{\bibinfo{person}{Angelika Bikner-Ahsbahs}, \bibinfo{person}{Christine Knipping}, {and} \bibinfo{person}{Norma Presmeg}} (Eds.). \bibinfo{publisher}{Springer Netherlands}, \bibinfo{address}{Dordrecht}, \bibinfo{pages}{365--380}.
\newblock


\bibitem[Melotti and de~Moraes(2019)]%
        {Melotti2019-rt}
\bibfield{author}{\bibinfo{person}{A~S Melotti} {and} \bibinfo{person}{C~H~V de Moraes}.} \bibinfo{year}{2019}\natexlab{}.
\newblock \showarticletitle{Evolving Roguelike Dungeons With Deluged Novelty Search Local Competition}.
\newblock \bibinfo{journal}{\emph{IEEE Trans. Comput. Intell. AI Games}} \bibinfo{volume}{11}, \bibinfo{number}{2} (\bibinfo{date}{June} \bibinfo{year}{2019}), \bibinfo{pages}{173--182}.
\newblock


\bibitem[Mo et~al\mbox{.}(2024)]%
        {Mo2024-ku}
\bibfield{author}{\bibinfo{person}{George Mo}, \bibinfo{person}{John Dudley}, \bibinfo{person}{Liwei Chan}, \bibinfo{person}{Yi-Chi Liao}, \bibinfo{person}{Antti Oulasvirta}, {and} \bibinfo{person}{Per~Ola Kristensson}.} \bibinfo{year}{2024}\natexlab{}.
\newblock \showarticletitle{Cooperative multi-objective Bayesian design optimization}.
\newblock \bibinfo{journal}{\emph{ACM Trans. Interact. Intell. Syst.}} \bibinfo{volume}{14}, \bibinfo{number}{2} (\bibinfo{date}{June} \bibinfo{year}{2024}), \bibinfo{pages}{1--28}.
\newblock


\bibitem[Mouret and Clune(2015)]%
        {Mouret2015-kz}
\bibfield{author}{\bibinfo{person}{Jean-Baptiste Mouret} {and} \bibinfo{person}{Jeff Clune}.} \bibinfo{year}{2015}\natexlab{}.
\newblock \showarticletitle{Illuminating search spaces by mapping elites}.
\newblock \bibinfo{journal}{\emph{arXiv preprint}} (\bibinfo{date}{April} \bibinfo{year}{2015}).
\newblock
\showeprint[arxiv]{1504.04909}~[cs.AI]


\bibitem[Ngoon et~al\mbox{.}(2018)]%
        {Ngoon2018-iu}
\bibfield{author}{\bibinfo{person}{Tricia~J Ngoon}, \bibinfo{person}{C~Ailie Fraser}, \bibinfo{person}{Ariel~S Weingarten}, \bibinfo{person}{Mira Dontcheva}, {and} \bibinfo{person}{Scott Klemmer}.} \bibinfo{year}{2018}\natexlab{}.
\newblock \showarticletitle{Interactive guidance techniques for improving creative feedback}. In \bibinfo{booktitle}{\emph{Proceedings of the 2018 CHI Conference on Human Factors in Computing Systems}}. \bibinfo{publisher}{ACM}, \bibinfo{address}{New York, NY, USA}.
\newblock


\bibitem[Ploin et~al\mbox{.}(2022)]%
        {Ploin2022-ml}
\bibfield{author}{\bibinfo{person}{A Ploin}, \bibinfo{person}{R Eynon}, \bibinfo{person}{Hjorth I.}, {and} \bibinfo{person}{M~A Osborne}.} \bibinfo{year}{2022}\natexlab{}.
\newblock \bibinfo{booktitle}{\emph{{AI} and the Arts: How Machine Learning is Changing Artistic Work. Report from the Creative Algorithmic Intelligence Research Project}}.
\newblock \bibinfo{publisher}{Oxford Internet Institute, University of Oxford, UK}.
\newblock


\bibitem[Pugh et~al\mbox{.}(2015)]%
        {Pugh2015-ra}
\bibfield{author}{\bibinfo{person}{Justin~K Pugh}, \bibinfo{person}{L~B Soros}, \bibinfo{person}{Paul~A Szerlip}, {and} \bibinfo{person}{Kenneth~O Stanley}.} \bibinfo{year}{2015}\natexlab{}.
\newblock \showarticletitle{Confronting the Challenge of Quality Diversity}. In \bibinfo{booktitle}{\emph{Proceedings of the 2015 Annual Conference on Genetic and Evolutionary Computation}}. \bibinfo{publisher}{ACM}, \bibinfo{pages}{967--974}.
\newblock


\bibitem[Ruela and Valdivia~Delgado(2018)]%
        {Ruela2018-ah}
\bibfield{author}{\bibinfo{person}{A~S Ruela} {and} \bibinfo{person}{K Valdivia~Delgado}.} \bibinfo{year}{2018}\natexlab{}.
\newblock \showarticletitle{{Scale-Free} Evolutionary Level Generation}. In \bibinfo{booktitle}{\emph{2018 {IEEE} Conference on Computational Intelligence and Games ({CIG})}}. \bibinfo{pages}{1--8}.
\newblock


\bibitem[Secretan et~al\mbox{.}(2011)]%
        {Secretan2011-xq}
\bibfield{author}{\bibinfo{person}{Jimmy Secretan}, \bibinfo{person}{Nicholas Beato}, \bibinfo{person}{David~B D'Ambrosio}, \bibinfo{person}{Adelein Rodriguez}, \bibinfo{person}{Adam Campbell}, \bibinfo{person}{Jeremiah~T Folsom-Kovarik}, {and} \bibinfo{person}{Kenneth~O Stanley}.} \bibinfo{year}{2011}\natexlab{}.
\newblock \showarticletitle{Picbreeder: a case study in collaborative evolutionary exploration of design space}.
\newblock \bibinfo{journal}{\emph{Evol. Comput.}} \bibinfo{volume}{19}, \bibinfo{number}{3} (\bibinfo{date}{May} \bibinfo{year}{2011}), \bibinfo{pages}{373--403}.
\newblock


\bibitem[Sidner et~al\mbox{.}(2005)]%
        {Sidner2005-nu}
\bibfield{author}{\bibinfo{person}{Candace~L Sidner}, \bibinfo{person}{Christopher Lee}, \bibinfo{person}{Cory~D Kidd}, \bibinfo{person}{Neal Lesh}, {and} \bibinfo{person}{Charles Rich}.} \bibinfo{year}{2005}\natexlab{}.
\newblock \showarticletitle{Explorations in engagement for humans and robots}.
\newblock \bibinfo{journal}{\emph{Artif. Intell.}} \bibinfo{volume}{166}, \bibinfo{number}{1-2} (\bibinfo{date}{Aug.} \bibinfo{year}{2005}), \bibinfo{pages}{140--164}.
\newblock


\bibitem[Swearngin et~al\mbox{.}(2020)]%
        {Swearngin2020-xt}
\bibfield{author}{\bibinfo{person}{Amanda Swearngin}, \bibinfo{person}{Chenglong Wang}, \bibinfo{person}{Alannah Oleson}, \bibinfo{person}{James Fogarty}, {and} \bibinfo{person}{Amy~J Ko}.} \bibinfo{year}{2020}\natexlab{}.
\newblock \showarticletitle{Scout: Rapid exploration of interface layout alternatives through High-Level design constraints}. In \bibinfo{booktitle}{\emph{Proceedings of the 2020 CHI Conference on Human Factors in Computing Systems}}. \bibinfo{publisher}{ACM}, \bibinfo{address}{New York, NY, USA}.
\newblock


\bibitem[Vermeesch(2012)]%
        {Vermeesch_2012}
\bibfield{author}{\bibinfo{person}{Pieter Vermeesch}.} \bibinfo{year}{2012}\natexlab{}.
\newblock \showarticletitle{On the visualisation of detrital age distributions}.
\newblock \bibinfo{journal}{\emph{Chemical Geology}}  \bibinfo{volume}{312–313} (\bibinfo{date}{June} \bibinfo{year}{2012}), \bibinfo{pages}{190–194}.
\newblock
\showISSN{0009-2541}
\urldef\tempurl%
\url{https://doi.org/10.1016/j.chemgeo.2012.04.021}
\showDOI{\tempurl}


\bibitem[Vincalek et~al\mbox{.}(2021a)]%
        {Vincalek2021-lz2}
\bibfield{author}{\bibinfo{person}{Jakub Vincalek}, \bibinfo{person}{Sean Walton}, {and} \bibinfo{person}{Ben Evans}.} \bibinfo{year}{2021}\natexlab{a}.
\newblock \showarticletitle{It's the Journey Not the Destination: Building Genetic Algorithms Practitioners Can Trust}. In \bibinfo{booktitle}{\emph{2021 Genetic and Evolutionary Computation Conference Companion ({GECCO} '21 Companion)}}. \bibinfo{publisher}{ACM}, \bibinfo{address}{New York, NY, USA}.
\newblock


\bibitem[Vincalek et~al\mbox{.}(2021b)]%
        {Vincalek2021-lz}
\bibfield{author}{\bibinfo{person}{Jakub Vincalek}, \bibinfo{person}{Sean Walton}, {and} \bibinfo{person}{Ben Evans}.} \bibinfo{year}{2021}\natexlab{b}.
\newblock \showarticletitle{A user-centered approach to evolutionary algorithms and their use in industry}.
\newblock \bibinfo{journal}{\emph{Cogent Eng.}} \bibinfo{volume}{8}, \bibinfo{number}{1} (\bibinfo{date}{Jan.} \bibinfo{year}{2021}).
\newblock


\bibitem[von Rymon~Lipinski et~al\mbox{.}(2019)]%
        {Von_Rymon_Lipinski2019-tk}
\bibfield{author}{\bibinfo{person}{B von Rymon~Lipinski}, \bibinfo{person}{S Seibt}, \bibinfo{person}{J Roth}, {and} \bibinfo{person}{D Ab{\'e}}.} \bibinfo{year}{2019}\natexlab{}.
\newblock \showarticletitle{Level Graph -- Incremental Procedural Generation of Indoor Levels using Minimum Spanning Trees}. In \bibinfo{booktitle}{\emph{2019 {IEEE} Conference on Games ({CoG})}}. \bibinfo{pages}{1--7}.
\newblock


\bibitem[Walton et~al\mbox{.}(2021)]%
        {Walton2021-vi}
\bibfield{author}{\bibinfo{person}{S Walton}, \bibinfo{person}{A Rahat}, {and} \bibinfo{person}{J Stovold}.} \bibinfo{year}{2021}\natexlab{}.
\newblock \showarticletitle{Evaluating {Mixed-Initiative} Procedural Level Design Tools using a {Triple-Blind} {Mixed-Method} User Study}.
\newblock \bibinfo{journal}{\emph{IEEE Trans. Comput. Intell. AI Games}} (\bibinfo{year}{2021}).
\newblock


\bibitem[Yan et~al\mbox{.}(2017)]%
        {Yan2017-au}
\bibfield{author}{\bibinfo{person}{Shuo Yan}, \bibinfo{person}{Gangyi Ding}, \bibinfo{person}{Hongsong Li}, \bibinfo{person}{Ningxiao Sun}, \bibinfo{person}{Zheng Guan}, \bibinfo{person}{Yufeng Wu}, \bibinfo{person}{Longfei Zhang}, {and} \bibinfo{person}{Tianyu Huang}.} \bibinfo{year}{2017}\natexlab{}.
\newblock \showarticletitle{Exploring audience response in performing arts with a brain-Adaptive Digital Performance system}.
\newblock \bibinfo{journal}{\emph{ACM Trans. Interact. Intell. Syst.}} \bibinfo{volume}{7}, \bibinfo{number}{4} (\bibinfo{date}{Dec.} \bibinfo{year}{2017}), \bibinfo{pages}{1--28}.
\newblock


\bibitem[Yannakakis et~al\mbox{.}(2014)]%
        {Yannakakis2014-rt}
\bibfield{author}{\bibinfo{person}{Georgios~N Yannakakis}, \bibinfo{person}{Antonios Liapis}, {and} \bibinfo{person}{Constantine Alexopoulos}.} \bibinfo{year}{2014}\natexlab{}.
\newblock \showarticletitle{Mixed-initiative co-creativity}. In \bibinfo{booktitle}{\emph{9th International Conference on the Foundations of Digital Games}} (Fort Lauderdale). \bibinfo{publisher}{Foundations of Digital Games}.
\newblock


\bibitem[Zhu et~al\mbox{.}(2018)]%
        {zhu2018explainable}
\bibfield{author}{\bibinfo{person}{Jichen Zhu}, \bibinfo{person}{Antonios Liapis}, \bibinfo{person}{Sebastian Risi}, \bibinfo{person}{Rafael Bidarra}, {and} \bibinfo{person}{G~Michael Youngblood}.} \bibinfo{year}{2018}\natexlab{}.
\newblock \showarticletitle{Explainable AI for designers: A human-centered perspective on mixed-initiative co-creation}. In \bibinfo{booktitle}{\emph{2018 IEEE Conference on Computational Intelligence and Games (CIG)}}. IEEE, \bibinfo{pages}{1--8}.
\newblock


\end{thebibliography}

%%
%% If your work has an appendix, this is the place to put it.

%\section{Author Statement}
%This is original work which has not been published elsewhere. It significantly refines and expands the technique we introduced for evaluating mixed-initiative tools\cite{Walton2021-vi}. This original work is carried out at a much larger scale with an application more generalisable to fields where human-AI collaborative work is being adopted. Our findings in this paper are more statistically powerful and have significant repercussions for researchers and practitioners in this field.

\end{document}